\documentclass[%
 amsmath,amssymb,
 aps, physrev,
 twocolumn,
 nofootinbib
]{revtex4-2}

\usepackage{graphicx}
\usepackage{dcolumn}
\usepackage{bm}
\usepackage{amsfonts,amssymb,amsmath,amsthm}
\usepackage{subcaption}
\usepackage{enumitem}
\usepackage[ruled,vlined,linesnumbered]{algorithm2e}

\theoremstyle{definition}

\theoremstyle{theorem}
\newtheorem{theorem}{Theorem}

\theoremstyle{proposition}
\newtheorem{proposition}{Proposition}

\theoremstyle{corollary}
\newtheorem{corollary}{Corollary}

\theoremstyle{remark}
\newtheorem{remark}{Remark}

\theoremstyle{assumption}
\newtheorem{assumption}{Assumption}

\theoremstyle{example}
\newtheorem{example}{Example}

\begin{document}

\preprint{APS/123-QED}

\title{\textbf{Efficiently and Reliably Measuring Information Processing Capacity in Dynamical Systems via Kernels} 
}%


\author{Alessio Benavoli}
\email{alessio.benavoli@tcd.ie}
\altaffiliation[]{School of Computer Science and Statistics, Trinity College Dublin. Trinity Quantum Alliance, Unit 16, Trinity Technology and Enterprise Centre, Pearse Street, Dublin 2, Ireland}
\author{Felix Binder}
\altaffiliation[]{School of Physics. Trinity Quantum Alliance, Unit 16, Trinity Technology and Enterprise Centre, Pearse Street, Dublin 2, Ireland}
\author{Cangxiong Chen}
\altaffiliation[]{School of Computer Science and Statistics, Trinity College Dublin. }
\affiliation{Trinity College Dublin, Dublin, Ireland.}


\date{\today}

\begin{abstract}
Information Processing Capacity (IPC) is a powerful system-agnostic metric for quantifying a dynamical system's computational capability and is widely used in quantum reservoir computing. However, standard IPC suffers from severe computational limitations. Computing IPC involves estimating how well the system can reconstruct, via linear regression, an (in principle) infinite family of orthogonal polynomial functions of delayed inputs. In practice, this evaluation must be made finite, so we need only to consider polynomials up to a chosen maximum degree. Moreover, the computational complexity for computing IPC  grows combinatorially with the considered maximum degree and delay.
Here, we introduce Kernel-IPC (KIPC), a scalable framework that uses kernel methods to compute the total IPC without explicitly enumerating basis functions. This allows us to account for the full (infinite) basis (without degree truncation). We also provide a null-hypothesis test for statistically comparing the KIPC of two dynamical systems. We validate KIPC on an echo state network and demonstrate its utility for quantum reservoir computing on a disordered spin chain.
\end{abstract}

\maketitle


\section{Introduction}
The \textit{Information Processing Capacity} (IPC) is a system-agnostic metric to quantify the ability of a dynamical system to perform
arbitrary transformations on an input signal. Originally proposed in \cite{dambre2012information}, the evaluation of the processing capacity is based on discrete inputs and discrete system
states. The idea is brilliant and simple. Assuming the dynamical system has the fading memory property, \cite{dambre2012information} shows that the input-output map belongs to a fading memory Hilbert space. Therefore, since any function in this space can be represented as a linear combination of its basis functions, a way to evaluate the IPC of a system is to assess its ability to reconstruct elements of this basis via linear regression from its internal states. Thus, the IPC can be
interpreted as the total number of linearly independent functions of the input signal that the system can compute.
This provides a powerful framework for comparing the IPC of different dynamical systems. As a system-agnostic metric, it offers a more principled alternative for quantifying informational capacity than evaluating performance across a suite of standard benchmark tasks (e.g., NARMA, delay line, etc.).

In \cite{dambre2012information} it is shown that the elements of this basis can be chosen as orthogonal polynomials (e.g., Legendre polynomials) of delayed input signals. The reason is related to the Volterra-series representation of fading-memory systems \cite{boyd2003fading}.
Extensions of IPC have been discussed in \cite{kubota2021unifying,hulser2023deriving,polloreno2026innovation}. In particular, \cite{kubota2021unifying} discusses how to compute IPC for time-variant systems, while the standard IPC can only be applied to time-invariant systems. Instead, \cite{hulser2023deriving} derives a connection between the IPC and the performance of the dynamical system acting as a reservoir to solve a specific benchmark task. Finally, \cite{polloreno2026innovation} introduces the innovation capacity to extend IPC to deal with a noisy reservoir.

IPC is particularly fundamental in \textit{quantum reservoir computing} (QRC). Because QRC is a rapidly evolving field actively exploring diverse hardware platforms \cite{nakajima2020physical,goldmann2021exploiting,koster2023limitations} and architectural designs, IPC provides a standardised task-independent framework to systematically evaluate and compare how different physical dynamics and reservoir configurations convert quantum resources into functional computational capacity. Indeed, a substantial body of literature on QRC relies extensively on IPC to evaluate and compare: (i) different QRC architectures; (ii) alternative strategies for embedding the input signal into quantum states; (iii) different read-out approaches; (iv) the impact of measurement back-action:
\cite{vinckier2015high,harkhoe2019task,ortin2020delay,martinez2021dynamical,ghosh2021quantum,mujal2021opportunities,nokkala2021gaussian,nishioka2022edge,xia2022reservoir,gotting2023exploring,martinez2023information,fry2023optimizing,kubota2023temporal,tsunegi2023information,palacios2024role,vcindrak2024enhancing,sannia2024dissipation,llodra2025quantum,jing2025quantum,vcindrak2025krylov,hahto2025smarter,keenan2026storage,hou2026high}

IPC is powerful but also has some \textit{serious drawbacks}. The IPC cannot be computed for all elements of the basis of the Hilbert space of fading-memory functions, as this basis is infinite-dimensional. To keep the computation feasible, IPC must be truncated by setting a maximum degree and delay for the evaluating polynomials (but the number of such basis functions grows combinatorially with the polynomial degree and the maximum delay). Consequently, standard IPC calculations are practically limited to systems with lower nonlinearity and delays. This issue is further exacerbated by standard early-stopping heuristics, which halt calculations of IPC terms (which are summed to compute the total IPC) within a given degree once consecutive terms yield negligible capacity. Because the calculation of IPC terms is performed in an arbitrary order (typically graded-lexicographical order), early stopping prematurely skips later terms which, despite coming later in the lexicographical order, may be easier for the system to reconstruct and, therefore, contribute to the total IPC. Therefore, in papers using IPC, the computation of total IPC is often accompanied by a \textit{warning} stating that the reported total-IPC may be underestimated due to truncation, see for instance \cite[Appendix B]{hahto2025smarter}.

In this paper, we address this issue by showing how kernels can be used to compute the total IPC without explicitly enumerating all basis functions, and so avoiding the drawbacks discussed above. This enables us to compute the IPC without imposing an upper bound on the polynomial degree and, consequently, to capture all possible nonlinearities. As a result, we obtain an efficient and scalable approach for estimating the IPC of highly nonlinear dynamical systems, which we call Kernel-IPC (KIPC). Additionally, we provide a way to statistically compare the KIPC of two dynamical systems using a single null-hypothesis test. This solves another prominent issue with the standard IPC, which instead requires a separate hypothesis test for each element of the basis function, which can lead to a high false discovery rate due to spurious correlations. We demonstrated these improvements in practice using an Echo State Network as a test-bed dynamical system. Finally, we showed how the efficiency and reliability of KIPC can be leveraged to drive research in QRC, focusing specifically on a disordered quantum spin chain reservoir.

\section{Background}

\subsection{Information Processing Capacity}
\label{sec:ipc}
Consider a discrete-time $n_s$-dimensional dynamical system $\mathcal{S}$. We use $t$ to denote the time index (with $t=0,1,2,\dots,n_T$), ${\bf x}_t \in \mathbb{R}^{n_s}$ the state at time $t$, $u_t \in \mathbb{R}$ the input signal at time $t$ and $\hat{y}_t \in \mathbb{R}$ the output signal at time $t$. 
Additionally, we use the notation without subscript, e.g., $u$, to generically refer to the signal. The dynamical system $\mathcal{S}$ we consider is as follows:
\begin{equation}
\label{eq:nonlindyn}
\mathcal{S}:\left\{\begin{array}{rl}
{\bf x}_{t+1}  &= f \left({\bf x}_t ,u_{t+1} \right) \\
\hat{y}_t &= \hat{\bf w}^\top {\bf x}_t + \hat{w}_0
\end{array}\right.
\end{equation}
where $\hat{\mathbf{w}} \in \mathbb{R}^{n_s}$ and $\hat{w}_0 \in \mathbb{R}$ are parameters. These parameters are tunable. We will refer to \eqref{eq:nonlindyn} as a reservoir computing system if $\mathcal{S}$ has the \textit{fading memory property} \cite{boyd2003fading}: an input–output system $\mathcal{S}$  is said to exhibit fading memory if two input signals $u,v$
which are close in the recent past, but not necessarily
close in the remote past, yield present outputs which are
close.

The input signal $u$ is assumed to be a zero-mean, i.i.d. sequence. We use the vector ${\bf u}=[u_1,u_2,\dots]^\top$ to denote the input sequence. The total \textit{information processing capacity} \cite{dambre2012information} of $\mathcal{S}$ is defined as:
\begin{equation}
\label{eq:singIPC}
\begin{aligned}
\mathrm{C}^\mathrm{tot}(\mathcal{S},\mathcal{B}, u) &:= \sum_{d=1}^{\infty} \mathrm{C}^d(\mathcal{S}, \mathcal{B}_d, u), ~\text{ with } \\
\mathrm{C}^d(\mathcal{S}, \mathcal{B}_d, u)  &:= \sum_{ {\bf b} \in \mathcal{B}_d}  \mathrm{C}^d(\mathcal{S}, {\bf b}, u)
\end{aligned}
\end{equation}
where $\mathcal{B}$ is a basis of the Hilbert space of the fading memory functions \cite{dambre2012information}. A typical choice for $\mathcal{B}$ are orthogonal polynomials of the input signal. In this case, we denote with $\mathcal{B}_d$ the set of all polynomials of degree $d$ in this basis function and $\bf b$ is a generic term of this basis function. Each term is a polynomial of degree $d$ of  delayed input signals; we will provide an example in the next paragraph. The (single) information processing capacity (IPC) term is then defined as:
\begin{equation}
\label{eq:IPCterm}
\begin{aligned}
 \mathrm{C}^d(\mathcal{S}, {\bf b}, u):= 1-\min_{\hat{\bf w},\hat{w}_0} \frac{{\sum\limits_{t=1}^{n_T}(b_t- \hat{\bf w}^\top {\bf x}_t - \hat{w}_0)^2}}{{\bf b}^\top {\bf b}},
\end{aligned}
\end{equation}
which expresses the capacity of $\mathcal{S}$ of reconstructing an individual element $\bf b$ of the basis $\mathcal{B}$ from its states ${\bf x}_t$  via linear estimation (with linear regression parameters $\hat{\bf w},\hat{w}_0$). Since the bias term $\hat{w}_0$ is explicitly included in the linear regression model, the  columns of ${\bf x}_t$ are generally assumed to be centered to zero mean \cite{dambre2012information}. Accordingly, we assume zero-mean centered ${\bf x}_t$ for the rest of this work.
A zero reconstruction error for the element ${\bf b}$ leads to a maximum capacity of  $\mathrm{C}^d(\mathcal{S}, {\bf b}, u)=1$.

\citet{dambre2012information} proved the following  asymptotic ($n_T \rightarrow \infty$) bound:
\begin{equation}
\label{eq:totIPCbounds}
 0\leq \mathrm{C}^\mathrm{tot}(\mathcal{S},\mathcal{B}, u)  \leq n_s,
\end{equation}
where $n_s$ is the dimension of the state ${\bf x}_t$. Therefore, the total capacity of a reservoir cannot exceed the number of its states.

We now describe how these basis functions in $\mathcal{B}_d$ are constructed and summarise the main issues of this method.

\paragraph{Constructing an orthogonal basis}
 Let us denote the $\tau$-delayed input sequence with ${u}^{(\tau)}=[u_{\tau+1},u_{\tau+2},\dots]^\top$ for $\tau=1,2,\dots$ and let $\mathcal{P}_d(u^{(\tau)})$ denote the evaluation of a polynomial of degree $d$ elementwise on the elements of the input sequence, i.e.,  $\mathcal{P}_d(u^{(\tau)})=[\mathcal{P}_d(u_{\tau+1}),\mathcal{P}_d(u_{\tau+2}),\dots]^\top$. Then the basis $\mathcal{B}_d$ is constructed as follows. The first-order polynomials are 
$\mathcal{B}_1=\{\mathcal{P}_1(u^{(1)}),~\mathcal{P}_1(u^{(2)}),~\mathcal{P}_1(u^{(3)}),~\mathcal{P}_1(u^{(4)}),\dots\}$, 
and the second-order are
\begin{equation}
    \label{eq:B2}
    \begin{aligned}
    \mathcal{B}_2=\{&\mathcal{P}_2(u^{(1)}),~\mathcal{P}_2(u^{(2)}),~\mathcal{P}_2(u^{(3)}),\mathcal{P}_2(u^{(4)}),\dots\\\
    &\mathcal{P}_1(u^{(1)}) \mathcal{P}_1(u^{(2)}),\mathcal{P}_1(u^{(1)}) \mathcal{P}_1(u^{(3)}),\mathcal{P}_1(u^{(1)}) \mathcal{P}_1(u^{(4)})\\
        &\mathcal{P}_1(u^{(2)}) \mathcal{P}_1(u^{(3)}),\mathcal{P}_1(u^{(2)}) \mathcal{P}_1(u^{(4)}),\dots\}.
\end{aligned}
\end{equation}
Note that, $\mathcal{B}_2$ also includes the product of polynomials of degree $1$. $\mathcal{B}_3$ is equal to:
\begin{equation}
    \label{eq:B2}
    \begin{aligned}
    &\mathcal{B}_3=\{\mathcal{P}_3(u^{(1)}),~\mathcal{P}_3(u^{(2)}),~\mathcal{P}_3(u^{(3)}),~\mathcal{P}_3(u^{(4)}),\dots\\
    &\mathcal{P}_2(u^{(1)}) \mathcal{P}_1(u^{(2)}),\mathcal{P}_2(u^{(1)}) \mathcal{P}_1(u^{(3)}),\ldots\\
            &\mathcal{P}_1(u^{(1)}) \mathcal{P}_1(u^{(2)}) \mathcal{P}_1(u^{(3)}),\mathcal{P}_2(u^{(1)}) \mathcal{P}_1(u^{(3)}) \mathcal{P}_1(u^{(4)}),\ldots\},\end{aligned}
\end{equation}
and it includes all polynomials of degree $3$, comprising the products of polynomials of degree two and degree one, and the product of  polynomials of degree one.
 
 If each element of the input sequence $u_i$ is generated uniformly at random in $[-1,1]$, then by choosing Legendre polynomials, $\mathcal{P}_d=\mathcal{L}_d$, the collection of polynomials $\mathcal{B}_1\cup\mathcal{B}_2\cup \mathcal{B}_3 \cup \dots$ forms an orthonormal basis of the Hilbert space of fading memory functions. Instead, if $u_i \sim N(0,1/2)$ (Gaussian distributed), then one can choose Hermite polynomials, $\mathcal{P}_d=\mathcal{H}_d$, to form an orthonormal basis.

For practical numerical evaluation of the IPC in \eqref{eq:IPCterm}, a cutoff \cite{dambre2012information} in the length of the input sequence and a maximum delay $\tau_{max}$ must be defined, since the IPC cannot be computed for all elements of the basis of the Hilbert space of fading memory functions, as the basis is infinite-dimensional. For instance, by choosing $\tau_{max}=3$,  $\mathcal{B}_1$ will only have three terms $\{\mathcal{P}_1(u^{(1)}),~\mathcal{P}_1(u^{(2)}),~\mathcal{P}_1(u^{(3)})\}$, with $$
\begin{aligned}
    &\mathcal{P}_1(u^{(\tau)})\\
    &=[\overbrace{\mathcal{P}_1(0),\dots,\mathcal{P}_1(0)}^{\tau},\mathcal{P}_1(u_{\tau+1}),\mathcal{P}_1(u_{\tau+2}),\dots,\mathcal{P}_1(u_{n_T})]^\top
    \end{aligned}
    $$ 
    for $\tau \leq 3$,     
    where we have assumed a fixed length $n_T$, so that ${\bf b}$ in \eqref{eq:IPCterm} belongs to $\mathbb{R}^{n_T}$.\\

\paragraph{Issues of the current method to compute IPC} 
These are the main issues of the current method to compute IPC as mentioned in the Introduction.
\begin{description}
    \item[Issue 1] The number of terms in each  $\mathcal{B}_d$ increases combinatorially with $d$ and $\tau_{max}$ as
    \begin{equation}
  |\mathcal{B}_d| = {\tau_{max} +d-1\choose d}.
    \end{equation}
    For instance, for  $\tau_{max}=20$ and $d=10$,
      $|\mathcal{B}_d|=20,030,010$.
    Therefore, calculating the IPC requires an exhaustive search over combinations of orthogonal polynomials and time delays, which causes the computational cost to grow combinatorially with increasing polynomial degree. In practice, this requires restricting the search to relatively low maximum degrees and short maximum delays.
   In \cite[Supplementary, Sec.~3.3]{dambre2012information}, this restriction is justified by the following implicit assumption.
    \begin{assumption}
    \label{ass:1}
The capacities of a dynamical system generally decrease
when the degree of the polynomial and the delay increase.
\end{assumption}
In \cite[Supplementary, Sec.~3.3]{dambre2012information}, the authors motivate Assumption \ref{ass:1} on the basis of two empirical observations: (1) in the systems considered experimentally, capacity generally decreases as the polynomial degree increases; and (2) for physical systems with fading memory, capacity decreases, on  average, as the delay increases.
\item[Issue 2] Consider two reservoirs $\mathcal{S}_a, \mathcal{S}_b$  with same maximum capacities, i.e., $\mathrm{C}^\mathrm{tot}(\mathcal{S}_a,\mathcal{B},u)=\mathrm{C}^\mathrm{tot}(\mathcal{S}_b,\mathcal{B},u)=n_s$. Additionally, assume that $\mathcal{S}_a$ exhibits only low-degree nonlinearities, while $\mathcal{S}_b$ is highly nonlinear (and so  it performs poorly on linear memory tasks, e.g., fitting the elements of the basis $\mathcal{B}_1$).
The total information capacity $\mathrm{C}^\mathrm{tot}$ cannot distinguish between these two reservoirs, although it would be natural (also in view of Assumption~\ref{ass:1}) to prefer the system $\mathcal{S}_a$.
\item[Issue 3] Due to the use of finite statistics (i.e., a finite time series of length $n_T$) which introduces spurious correlations, capacities cannot be estimated reliably. To mitigate the effects of these numerical errors, which affect the value of the total capacity when $\tau_{\max}$ and $d$ are large, \cite{dambre2012information} suggest performing hypothesis testing to remove IPC terms with small values. Under an independence assumption (which is, in fact, violated \cite{dambre2012information}), they show that the total capacity is a sum of $\chi^2$-distributed variables. 
 Consequently, they define a threshold $r$ via $P(\chi^2 > r) = p$ with $p = 0.0001$ and then use the threshold $r$ to eliminate all small IPC terms (assuming they are due to  spurious correlations). They then double this threshold to account for the fact that the total IPC is not the sum of independent $\chi^2$ terms. This procedure is heuristic and, more importantly, does not adjust for multiple testing (Bonferroni-like corrections) -- the number of tests becomes large when $\tau_{\max}$ and $d$ are large. This means that the total IPC will mostly of the case be inflated due to spurious correlations when  $\tau_{\max}$ and $d$ are large.
\item[Issue 4] In the computation of total IPC, the choice of the particular orthogonal basis of polynomial is determined
by the distribution of $u$, for example, Legendre polynomials are used if $u$ is uniformly distributed in $[-1,1]$. However, the distribution of $u$ may be non-standard and unknown and, therefore, the IPC is not optimally  calculable with the formulas \eqref{eq:singIPC}.
\end{description}
%

%

\section{Computing IPC through kernels}
This section presents the main contribution of the paper. Building on the background material, we show how kernels can be used to compute the total IPC without explicitly considering all basis functions and, therefore, avoiding the previously discussed drawbacks.
Consider the function
$$
 f(u_t) = \sum_{n=0}^d \gamma_n \mathcal{P}_n(u_t),
$$
which is a linear combination of polynomial functions
of degree $n$. Assuming that the coefficients satisfy $\gamma_n \sim \mathcal{N}(0,\sigma_n^2)$ independently for $n=0,\dots,d$, then $f$ is zero-mean Gaussian distributed with  covariance  (or kernel function)
$$
\mathbb{E}[f(u_t)f(u_{t'})] 
  = k_d(u_t,u_{t'}) 
  = \sum_{n=0}^d \sigma_n^2 \, \mathcal{P}_n(u_t)\mathcal{P}_n(u_{t'}),
$$
where $u_{t},u_{t'}$ are two inputs at time $t$ and, respectively, $t'$.
We are interested in the limit $d\to\infty$ for various systems of orthogonal polynomials and we want to choose $\sigma_n^2$ so that the limit
converges to a known (preferably simple) closed-form function.

This leads to the following general problem: for a given family of orthogonal polynomials $\{\mathcal{P}_n\}_{n\ge0}$, find a sequence of variances $\{\sigma_n^2\}_{n\ge0}$ such that the infinite series defining $k_{\infty}(u_t,u_{t'})$ converges    and can be identified with a familiar kernel. 
Fortunately, this problem has already been solved for a broad class of orthogonal polynomials, often in the context of \textit{mathematical physics}.

\begin{proposition}
For Hermite polynomials, assuming that $\sigma_n^2=\frac{\rho^n}{2^n n!}$, we have that the following infinite series converges:
\begin{align}
\nonumber
&\sum_{n=0}^{\infty} \frac{\rho^n}{2^n n!} \mathcal{H}_n(u_t) \mathcal{H}_n(u_{t'})  
\label{eq:Herm}
\\&= \tfrac{1}{\sqrt{1 - \rho^2}} \exp \left( \tfrac{-\rho^2 (u_t^2 + u_{t'}^2) + 2\rho u_t u_{t'}}{1 - \rho^2} \right)=:
K_{Herm}(u_t,u_{t'}),
\end{align}
with $u_t,u_{t'} \in \mathbb{R}$ and $\rho \in (0,1)$. 

For Laguerre polynomials, assuming that $\sigma_n^2=\tfrac{n! \rho^n}{\Gamma(n+\alpha+1)}$, we have that:
\begin{align}
\nonumber
&\sum_{n=0}^{\infty} \tfrac{n! \rho^n}{\Gamma(n+\alpha+1)} \mathcal{G}^{(\alpha)}_n(u) \mathcal{G}^{(\alpha)}_n(u_{t'}) 
\\ &= {\tfrac {1}{(uu_{t'}\rho)^{\alpha /2}(1-\rho)}}e^{-\tfrac{\rho (u+u_{t'})}{1-\rho}}I_{\alpha }\left({\tfrac {2{\sqrt {uu_{t'} \rho}}}{1-\rho}}\right)=:K_{Lag}(u,u_{t'}),
\end{align}
with $u,u_{t'} \in \mathbb{R}^+$, $\alpha>-1$ and $\rho \in (0,1)$, where $I_{\alpha}$ denotes the modified Bessel function of the first kind.

For Jacobi polynomials, assuming that $\sigma_n^2=\tfrac{n!(\alpha+\beta+1)_n \rho^{n}}{(\alpha+1)_n(\beta+1)_n}$, we have that: 
\begin{equation}
\label{eq:jac}
 \begin{aligned}&\sum _{n=0}^{\infty }\tfrac{n!(\alpha+\beta+1)_n \rho^{n}}{(\alpha+1)_n(\beta+1)_n} P_{n}^{(\alpha ,\beta )}(u_t )P_{n}^{(\alpha ,\beta )}(u_{t'} )\\
 &=(1+\rho)^{-\alpha-\beta-1}F_{4}\Big({\tfrac {\alpha +\beta +1}{2}},{\tfrac {\alpha +\beta +2}{2}};\alpha +1,\beta +1;
 \\&\hspace{1cm}\tfrac{(1-u_t)(1-u_{t'})\rho}{(1+\rho)^2},\tfrac{(1+u_t)(1+u_{t'})\rho}{(1+\rho)^2}\Big)=:K_{Jac}(u_t,u_{t'})\end{aligned}
\end{equation}
with $u_t,u_{t'} \in [-1,1]$  ${\displaystyle \rho \in(0,1),\alpha +\beta >-1}$, where  ${\displaystyle (\cdot)_{n}}$ is Pochhammer's symbol (for the rising factorial) and ${\displaystyle F_{4}}$ is Appel's hypergeometric function.
\end{proposition}
The first kernel is known as \textit{Mehler kernel} \cite{mehler1866ueber}. It also corresponds to the Green's function of the Hamiltonian for the quantum harmonic oscillator and it is also related to the square-exponential kernel \cite[Sec.\ 4.3]{williams2006gaussian}. The second one for Laguerre polynomials is known as \textit{Hardy-Hille formula} \cite{hardy1932summation}. 
The last one is known as \textit{Bailey’s formula} \cite{Srivastava1987ANO}.  We recall that the \textit{Jacobi polynomials} ${\displaystyle P_{n}^{(\alpha ,\beta )}(z)}$ are a class of   polynomials, which  are orthogonal with respect to the weight ${\displaystyle (1-z)^{\alpha }(1+z)^{\beta }}$ on the interval 
${\displaystyle [-1,1]}$ and, that, the Chebyshev and Legendre,  (and others) polynomials, are special cases of the Jacobi polynomials. We then exploit the following result.

Note that, if $u_t,u_{t'}  \sim N(0,1/2)$, then 
$$
\int K_{Herm}(u_t,u_{t'})
\mathcal{H}_n(u_{t'})   N(u_{t'};0,1/2) du_{t'}=
\rho^n \mathcal{H}_n(u_t), 
$$
and, therefore, the kernel in \eqref{eq:Herm} has eigenfunctions  $\mathcal{H}_n$ and eigenvalues $\rho^n$. In the rest of the paper, we will focus on Legendre polynomials.
\begin{corollary}
\label{co:1}
For $\alpha,\beta = 0$, Jacobi polynomials reduce to Legendre polynomials, and Bailey’s formula \eqref{eq:jac} can be rewritten as
\begin{equation}
\label{eq:leg}
\begin{aligned}
&\sum_{n=0}^{\infty} \rho^n\mathcal{L}_n(u_{t})\mathcal{L}_n(u_{t'})  \\
&= \frac{2}{\pi} \tfrac{R\left( \sqrt{\frac{2t_2(u_{t},u_{t'})}{t_1(u_{t},u_{t'}) - t_2(u_{t},u_{t'})}}\right)}{\sqrt{t_1(u_{t},u_{t'})-t_2(u_{t},u_{t'})}}:=K_{Leg}(u_t,u_{t'}),
\end{aligned}
\end{equation}
where $t_1(u_{t},u_{t'}) = 1 - 2u_{t}u_{t'} \rho + \rho^2$, $t_2(u_{t},u_{t'}) = 2\rho\sqrt{1 - u_{t}^2} \sqrt{1 - u_{t'}^2}$  and $R(z)=\int_{0}^{\pi/2} \tfrac{d \theta}{\sqrt{1-z^2 \sin^2(\theta)}}$  is the complete elliptic integral of the first kind.
\end{corollary}
The proof of this corollary and next theorems is in Appendix \ref{app:proofs}. This kernel was derived 
in \cite{HANRAHAN2023105921}  to find  sharp estimates for the Jacobi heat kernels, in Corollary \ref{co:1} instead we show its derivation from Bailey’s formula.
Note that, if $u,u_{t'}  \sim \text{Unif}(-1,1)$, then 
\begin{equation}
\label{eq:eigenleg}
   \int K_{Leg}(u,u_{t'})
\mathcal{L}_n(u_{t'})   \text{Unif}(u_{t'};-1,1) du_{t'}=
\tfrac{\rho^n}{2n+1} \mathcal{L}_n(u),  
\end{equation}
and, therefore, the kernel in \eqref{eq:leg} has eigenfunctions  $\mathcal{L}_n$ and eigenvalues $\rho^n/(2n+1)$.

To extend the previous kernel from a scalar to an $m$-dimensional vector of delayed input-signals, that is, ${\bf u}_t := [u_{t-1}, u_{t-2}, \dots, u_{t-m}]^\top \in \mathbb{R}^m$, we can use the tensor product. Given  a kernel $k: \mathbb{R} \times \mathbb{R} \rightarrow \mathbb{R}$, the tensor product kernel, $k_{\otimes}: \mathbb{R}^m \times \mathbb{R} ^m\rightarrow \mathbb{R}$,  is defined as
\begin{equation}
\label{eq:kernelindprod}
    K_{\otimes}({\bf u}_{t},{\bf u}_{t'}) := \prod_{\tau=1}^{m}  K(u_{t-\tau},{u}_{t'-\tau}),
\end{equation}
that is, by considering the product of $m$ kernels each one defined on a specific element of the vector ${\bf u}_{t}$. The basis functions of the tensor product kernel are the tensor products of the individual basis functions.
Therefore, if we build a tensor product kernel from the infinite dimensional kernels, such as for instance $K_{Leg}(u_t,u_{t'})$, the tensor product kernel will include the product of all  polynomials of any degree of the elements of the vector  ${\bf u}_{t} \in \mathbb{R}^m$.
This leads to the general expression
\begin{equation}
\label{eq:KproductP}
\begin{aligned}
 K_{\otimes}({\bf u}_t, {\bf u}_{t'}) &= \prod_{\tau=1}^{m}    \sum_{n=0}^{\infty} \overline{\lambda}_n\overline{\mathcal{P}}_n(u_{t-\tau}) \overline{\mathcal{P}}_n(u_{t'-\tau})\\
 &=\sum\limits_{\boldsymbol{\iota} \in \mathbb{N}^m} \overline{\lambda}_{\boldsymbol{\iota}} \overline{\mathcal{P}}^{\otimes}_{\boldsymbol{\iota}}({\bf u}_t)  \overline{\mathcal{P}}^{\otimes}_{\boldsymbol{\iota}}({\bf u}_{t'})
 \end{aligned}
\end{equation}
where we use $\overline{\mathcal{P}}_n$ to denote generically an orthonormal polynomial of degree $n$ which has been scaled to have normalisation one (with respect to the corresponding probability measure), $\overline{\lambda}_i$ is the scaled eigenvalue (e.g., $\overline{\lambda}_i=\rho^i$ for Hermite and $\overline{\lambda}_i=\rho^i/(2 i +1)$ for Legendre) and  $\overline{\lambda}_{\boldsymbol{\iota}}$ and $\overline{\mathcal{P}}^{\otimes}_{\boldsymbol{\iota}}({\bf u}_t)$ denote the product of the eigenvalues and polynomials based on the signature $\boldsymbol{\iota}$.
For instance, for the Legendre case, assuming that $m=4$ and $\boldsymbol{\iota}=(2,1,1,3)$, we have that ${\lambda}_{(2,1,1,3)}=\frac{\rho^2}{5}\frac{\rho}{3}\frac{\rho}{3}\frac{\rho^3}{7}=\frac{\rho^7}{315}$ and $\overline{\mathcal{P}}^{\otimes}_{(2,1,1,3)}({\bf u}_{t})=\overline{\mathcal{P}}_2(u_{t-1})\overline{\mathcal{P}}_1(u_{t-2})\overline{\mathcal{P}}_1(u_{t-3})\overline{\mathcal{P}}_3(u_{t-4})$.

We will now derive the main result of the paper: the kernel-IPC (KIPC) metric.
\begin{theorem}
\label{th:1}
Let $X=[{\bf x}_1,\dots,{\bf x}_{n_T}]^\top \in \mathbb{R}^{n_T \times n_s}$ be the regressor matrix including the states of the  dynamical system $\mathcal{S}$  and $U=[{\bf u}_1,\dots,{\bf u}_{n_T}]^\top \in \mathbb{R}^{n_T \times \tau_{max}}$ the matrix of the delayed inputs. Then
\begin{equation}
\label{eq:KTPC}
\begin{aligned}
\lim_{n_T \rightarrow \infty} \frac{1}{n_T}Tr(\Pi K_{\otimes}(U,U))  &= \sum_{\boldsymbol{\iota} \in \mathbb{N}^m}  {\lambda}_{\boldsymbol{\iota}}\frac{(\mathcal{P}^{\otimes}_{\boldsymbol{\iota}}(U))^\top \Pi \mathcal{P}^{\otimes}_{\boldsymbol{\iota}}(U)}{(\mathcal{P}^{\otimes}_{\boldsymbol{\iota}}(U))^\top \mathcal{P}^{\otimes}_{\boldsymbol{\iota}}(U)}\\
&= \sum_{\boldsymbol{\iota} \in \mathbb{N}^m}  {\lambda}_{\boldsymbol{\iota}} IPC_{\boldsymbol{\iota}}
\end{aligned}
\end{equation}
where $\Pi=X (X^\top X)^{-1} X^\top$ is the projection matrix defined by $X$ and $IPC_{\boldsymbol{\iota}}$ is the expression \eqref{eq:IPCterm} for the element of the basis function ${\bf b}_{\boldsymbol{\iota}}$.
\end{theorem}
We have just shown that $\frac{1}{n_T}\operatorname{Tr}(\Pi K_{\otimes}(U,U))$ computes the total IPC, with two important differences from the standard definition in \eqref{eq:IPCterm}. First, the sum extends to infinity: we do not need to truncate it, so the kernel-total IPC considers polynomials of all possible degrees. Second, the weights $\lambda_{\boldsymbol{\iota}}$ differ: in the standard total IPC these weights are all equal to one (since practitioners cannot evaluate infinitely many IPC terms, the infinite sum must be truncated in practice), whereas in the kernel-total IPC we have $\lambda_{\boldsymbol{\iota}} < 1$ for $\boldsymbol{\iota} \neq \boldsymbol{0}$, with the eigenvalues decreasing as the degree increases (and the sum goes to infinity). The condition $\lambda_{\boldsymbol{\iota}} < 1$ is required to ensure that the series in \eqref{eq:KproductP} converges.

\begin{remark}
\label{rem:1}
The fact that we make the eigenvalues ${\lambda}_{\boldsymbol{\iota}}$ in \eqref{eq:KTPC} to decrease with increasing degree reflects the implicit standard practice for computing  IPC. In the standard computation of total IPC it is not possible to evaluate the capacities for all basis functions, which are infinite. Since one does not know in advance which functions the dynamical system can compute and, therefore, which basis functions to evaluate, the standard strategy \cite{dambre2012information} for computing the total IPC starts with polynomials of degree one and then increases the degree until the capacity falls below a given threshold.
In other words, this strategy assigns more importance to lower-degree than to higher-degree polynomials in the basis functions. In KIPC, this is exactly the importance that the weights ${\lambda}_{\boldsymbol{\iota}}$ capture and, therefore, in KIPC we make Assumption \ref{ass:1} explicit by introducing the weights ${\lambda}_{\boldsymbol{\iota}}$. Assumption 1 also states that the weights should decrease with the delay, we will discuss about this in Section \ref{sec:rho}.~\\
\end{remark}

Hereafter, we provide an useful asymptotic bound for KIPC.

\begin{theorem}
\label{th:2}
Let ${\lambda}_{0}, \dots, {\lambda}_{n_s}$ be the largest eigenvalues in the set $\{\lambda_{\boldsymbol{\iota}}\}_{\boldsymbol{\iota} \in \mathbb{N}^m}$ in  \eqref{eq:KproductP}, then 
KIPC is bounded by
\begin{equation}
\label{eq:asymlim}
\lim\limits_{n_T \rightarrow \infty} \frac{1}{n_T}Tr(\Pi K_{\otimes}(U,U))  \leq \sum_{n=0}^{n_s} {\lambda}_{n}.
\end{equation}
\end{theorem}
The asymptotic upper bound in \eqref{eq:asymlim} for KIPC is the analogue of \eqref{eq:totIPCbounds} for the standard total-IPC.


\begin{example}
\label{ex:1}
Assume $\tau_{max}=m = 8$ and consider the following two regression matrices:
$$
\begin{aligned}
X_1 &= [\mathcal{L}_1(U_{\cdot 1}),\, \mathcal{L}_2(U_{\cdot 1}),\, \mathcal{L}_2(U_{\cdot 1})\mathcal{L}_1(U_{\cdot 2})],\\
X_2 &= [\mathcal{L}_1(U_{\cdot 1}),\, \mathcal{L}_{50}(U_{\cdot 1}),\, \mathcal{L}_2(U_{\cdot 1})\mathcal{L}_1(U_{\cdot 2})],
\end{aligned}
$$
where $U_{\cdot i}$ denotes the $i$-th column of the matrix $U$ defined in Theorem \ref{th:1}. Since the regression matrices are constructed directly from Legendre polynomials, the theoretical standard total IPC satisfies
\[
\mathrm{C}^{\mathrm{tot}}(X_1, \mathcal{L}, u)=\mathrm{C}^{\mathrm{tot}}(X_2, \mathcal{L}, u) = 3,
\]
in the limit $n_T \rightarrow \infty$. This follows by \eqref{eq:totIPCbounds}.  However, due to the presence of the degree-50 polynomial $\mathcal{L}_{50}$ in $X_2$ and assuming the components of $X_2$ are unknown, systematically recovering $\mathcal{L}_{50}$ (and thus obtaining $\mathrm{C}^{\mathrm{tot}}(X_2, \mathcal{L}, u) \approx 3$) would require evaluating at least
$
\sum_{d = 1}^{50} {m + d - 1 \choose d} = 1,916,797,310
$
individual IPC terms. This cannot be obtained computationally and, as a result, the total IPC for $X_2$ \textbf{will be incorrectly estimated as 2} due to truncation.

With kernel IPC, we do not need to worry about that as it considers all the possible degrees. Theoretically, from \eqref{eq:eigenleg}, we have that
$\lim_{n_T \rightarrow \infty} \frac{1}{n_T}\operatorname{Tr}(\Pi_1 K_{\otimes}(U,U)) = \frac{\rho}{3}+\frac{\rho^2}{5}+\frac{\rho^2}{5}\frac{\rho}{3}$ and $
\lim_{n_T \rightarrow \infty} \frac{1}{n_T}\operatorname{Tr}(\Pi_2 K_{\otimes}(U,U)) = \frac{\rho}{3}+\frac{\rho^{50}}{101}+\frac{\rho^2}{5}\frac{\rho}{3}$, 
where $\Pi_i=X_i (X_i^\top X_i)^{-1} X_i^\top$ for $i=1,2$. These are asymptotic  values ($n_T \rightarrow \infty$). Hereafter, we provide a finite statistic example. We generated $X_1$ and $X_2$ with different $n_T$ and computed KIPC for $\rho=0.95$:

{\begin{center}
\begin{tabular}{|c|c|c|c|c|}
    & \multicolumn{4}{c|}{$n_T$}\\
$\frac{1}{n_T}\operatorname{Tr}(\Pi_i K_{\otimes}(U,U))$ & $10^4$ & $10^5$ & $3\cdot 10^5$ & $8\cdot 10^5$\\\hline
$i=1$ & 0.829 & 0.583 & 0.561 & 0.556\\
$i=2$ & 0.651 & 0.399 & 0.383 & 0.376\\
\end{tabular}
\end{center}}
The theoretical values are $\approx 0.5543$ (i=1) and $\approx 0.3745$ (i=2) and we can see that KIPC is converging to those values with increasing $n_T$. The computational time is 0.4s ($n_T=10^4$),  22s  ($n_T=10^5$), 173s ($n_T=3 \cdot 10^5$) and 1200s ($n_T=8 \cdot 10^5$) on a laptop with Nvidia GPU RTX PRO 3000 Blackwell.
\end{example}
From the numerical example above, it can be seen that, for finite $n_T$, the quantity $\tfrac{1}{n_T}\operatorname{Tr}(\Pi K_{\otimes}(U,U))$ yields an upward-biased estimate of KIPC. We now provide an unbiased estimator. Note that the quantity $\tfrac{1}{n_T}\operatorname{Tr}(\Pi K_{\otimes}(U,U))$ can be decomposed as
\begin{align}
\nonumber
\frac{1}{n_T}\operatorname{Tr}(\Pi K_{\otimes}(U,U))&=\frac{1}{n_T^2}\sum_{i=1}^{n_T} n_T\Pi_{ii} K_{\otimes}(U,U)_{ii}\\
&+\frac{1}{n_T^2}\sum_{i=1}^{n_T} \sum_{j=1,\,j \neq i}^{n_T} n_T \Pi_{ij} K_{\otimes}(U,U)_{ij},
\end{align}
which is therefore a V-statistic (see for instance \cite{boroskikh2020u}), since the function $\Pi_{ij} K_{\otimes}(U,U)_{ij}=\Pi_{ji} K_{\otimes}(U,U)_{ji}$ is symmetric.
V-statistics are biased (for finite $n_T$). We can obtain an unbiased estimator of KIPC (for finite $n_T$) by removing the diagonal term and dividing the second term by $n_T(n_T-1)$ (the number of terms in the sum) instead of $n_T^2$.
\begin{proposition}
\label{prop:U}
 For finite $n_T$, an unbiased estimator for KIPC is   
 \begin{align}
\label{eq:KIPCunb}
\widehat{\mathrm{KIPC}} &:=\frac{1}{n_T-1}\sum_{i=1}^{n_T} \sum_{j=1,\,j \neq i}^{n_T} \Pi_{ij} K_{\otimes}(U,U)_{ij}.
\end{align}
\end{proposition}
This corresponds to a \text{U-statistic}  (see for instance \cite{boroskikh2020u}).  
This expression also shows that, although the matrices $\Pi,K_{\otimes}(U, U)$ are of dimension $n_T \times n_T$. we can compute the KIPC without explicitly allocating these matrices in memory. Appendix \ref{app:alg} details the algorithm we use to compute KIPC.

\begin{remark}
The computational complexity of computing KIPC is $\mathcal{O}(m\,n_T^2)$, where $m=\tau_{\max}$. The dependence on $m$ arises from \eqref{eq:kernelindprod}.
\end{remark}

\begin{example}
\label{ex:2}
We repeat the numerical experiment in Example \ref{ex:1} by using the unbiased estimator in \eqref{eq:KIPCunb}.

{\begin{center}
\begin{tabular}{|c|c|c|c|}
    & \multicolumn{3}{c|}{$n_T$}\\
$\widehat{\mathrm{KIPC}}$ in Eq. \eqref{eq:KIPCunb} & $10^4$ & $10^5$ & $3\cdot 10^5$ \\\hline
$i=1$ & 0.552 & 0.554 & 0.554 \\
$i=2$ & 0.373 & 0.374 & 0.374 \\
\end{tabular}
\end{center}}
We recall that the theoretical values are approximately 0.5543 (i=1) and 0.3745 (i=2). We observe that the unbiased KIPC estimator converges to these values much more quickly, reaching near-accurate estimates already for $n_T=10^4$. The computational time is exactly the same as in Example \ref{ex:1}.
\end{example}



Finally, in standard IPC it is common to divide the total IPC in the linear versus nonlinear contribution. The linear contribution is computed only considering the elements of the basis $\mathcal{P}_1$ (degree one polynomials).
We can also separate the  linear versus nonlinear contribution in KIPC.
\begin{corollary}
\label{co:2}
Let us define the finite dimensional kernel
\begin{equation}
    K^{(lin)}_\otimes(U,U) :=\sum_{\tau=1}^{\tau_{max}} \lambda_{1,\tau}\mathcal{P}_1(u_{t-\tau})\mathcal{P}_1(u_{t'-\tau}),
\end{equation}
 where we use $\lambda_{1,\tau}$ to denote the weight corresponding to the degree one polynomial for the delay $\tau$ (in the Legendre case $\lambda_{1,\tau}=\rho/2$, see \eqref{eq:leg}).  
    The linear-component of $\widehat{\mathrm{KIPC}}$ is then 
    \begin{equation}
        \widehat{\mathrm{KIPC}}_{lin}=\frac{1}{n_T-1}\sum_{i=1}^{n_T} \sum_{j=1,\,j \neq i}^{n_T} \Pi_{ij} K^{(lin)}_{\otimes}(U,U)_{ij}
    \end{equation}
 and the nonlinear-component of  KIPC can be computed as the difference $\widehat{\mathrm{KIPC}}_{nonlin}=\widehat{\mathrm{KIPC}}-\widehat{\mathrm{KIPC}}_{lin}$.
\end{corollary}

\begin{example}
\label{ex:3}
We repeat the numerical experiment in Example \ref{ex:2} with $n_T=10^5$ to compute $\widehat{\mathrm{KIPC}}_{lin},\widehat{\mathrm{KIPC}}_{nonlin}$

{\begin{center}
\begin{tabular}{|c|c|c|}
    & \multicolumn{2}{c|}{$n_T=10^5$}\\
 & $\widehat{\mathrm{KIPC}}_{lin}$ & $\widehat{\mathrm{KIPC}}_{nonlin}$ \\\hline
$i=1$ & 0.316 & 0.238 \\
$i=2$ & 0.316 & 0.058 \\
\end{tabular}
\end{center}}
In both cases ($i=1,2$), the theoretical linear KIPC is $\rho/3 \approx 0.316$, because for both $X_1,X_2$ the only linear component is $\mathcal{L}_1(U_{\cdot 1})$.
\end{example}

We have seen that, through KIPC, we have addressed \textbf{Issue 1}, since we can compute the information capacity without imposing an upper bound on the polynomial degree and, consequently, we account for all possible nonlinearities. Moreover, this can be done efficiently  -- in a fraction of a second for $n_T=10^4$, as discussed in Example \ref{ex:1}.

Additionally, we have addressed \textbf{Issue 2} by automatically down-weighting higher-degree polynomials. This is also evident in Example~\ref{ex:1}, where two reservoirs with the same rank (the rank of $X_1,X_2$ is rank=$3$) have different capacities, with the second exhibiting a lower capacity  due to the higher-degree polynomial term $\mathcal{L}_{50}$.

\subsection{Hypothesis testing}
\label{sec:hypo}
We can use KIPC to statistically assess whether two reservoirs, $\mathcal{S}_a$ and $\mathcal{S}_b$, driven by the same input signal of length $n_T$, exhibit significantly different capacities. Because the $\widehat{\mathrm{KIPC}}$ estimator in \eqref{eq:KIPCunb} is unbiased, under the null hypothesis $H_0$ (i.e., $H_0$: $\mathcal{S}_a$ and $\mathcal{S}_b$ are equivalent), we have
\begin{align}
\label{eq:KIPCdelta}
\Delta&:=\frac{1}{n_T-1}\sum_{i=1}^{n_T} \sum_{j=1,\,j \neq i}^{n_T} (\Pi^{(a)}_{ij}- \Pi^{(b)}_{ij}) K_{\otimes}(U,U)_{ij},
\end{align}
is such that $E[\Delta]=0$, where the expectation is taken with respect to the distribution of the input signal.
Additionally, under the null-hypothesis, the two projection matrices $\Pi_a,\Pi_b$ are projecting on the same space and, therefore, the difference $\Pi_a - \Pi_b$ reflects purely random variations. Therefore, $\Pi_a - \Pi_b$  and $K_{\otimes}(U,U)$ can be assumed to be independent. Therefore, we can use a permutation-based approach to compute the distribution under $H_0$. This permutation-based approach is commonly used in kernel-based hypothesis testing, see for instance \cite{gretton2007kernel}.

Let $\pi$ be a permutation of the signal  $u$ and let $U_\pi$ be the corresponding matrix, define
\begin{align}
\label{eq:KIPCdelta1}
d_\pi&:=\frac{1}{n_T-1}\sum_{i=1}^{n_T} \sum_{j=1,\,j \neq i}^{n_T} (\Pi^{(a)}_{ij}- \Pi^{(b)}_{ij}) K_{\otimes}(U_\pi,U_\pi)_{ij},
\end{align}
Note that, $\Pi_a - \Pi_b$ is not permuted.
Then, we can approximate the distribution under $H_0$ by collecting the sample statistics $\mathcal{D}=\{d_\pi : \pi = 1, \dots, n_{\text{samples}}\}$ for $n_{\text{samples}}$ different random-permutations. Since the distribution  under $H_0$ is symmetric around zero (because $\Pi_a - \Pi_b$ can be positive or negative with equal probability), we can use the $\alpha/2$ and $1-\alpha/2$ percentiles of $\mathcal{D}$ to perform an $\alpha$-level two-sided $p$-value-based hypothesis test.

\begin{example}
We again consider $X_1,X_2$ in Example \ref{ex:1} and additionally
$$
\begin{aligned}
X_3 &= [\mathcal{L}_1(U_{\cdot 2}),\, \mathcal{L}_2(U_{\cdot 2}),\, \mathcal{L}_2(U_{\cdot 2})\mathcal{L}_1(U_{\cdot 3})],\\
X_4 &= [\mathcal{L}_1(U_{\cdot 2}),\, \mathcal{L}_{50}(U_{\cdot 2}),\, \mathcal{L}_2(U_{\cdot 2})\mathcal{L}_1(U_{\cdot 3})],
\end{aligned}
$$
It is evident that $X_1,X_3$ and $X_2,X_4$ have the same KIPC: $\approx 0.5543$ (i=1,3) and $\approx 0.3745$ (i=2,4).

We select $n_T=10^4$ and perform a hypothesis testing with $\alpha=0.05$ and $n_{samples}=300$ to test the difference in capacity between these four reservoirs. The result is shown in Figure \ref{fig:test}. It is evident that we cannot reject the null-hypothesis (with significance $\alpha$) for the comparisons $X_1,X_3$ and $X_2,X_4$. Conversely, we can reject the null (as expected) for the comparisons $X_1,X_2$ and $X_3,X_4$.

\begin{figure*}[hbt!]
  \centering
   \begin{subfigure}{.5\linewidth}
    \centering
    \includegraphics[width=\linewidth]{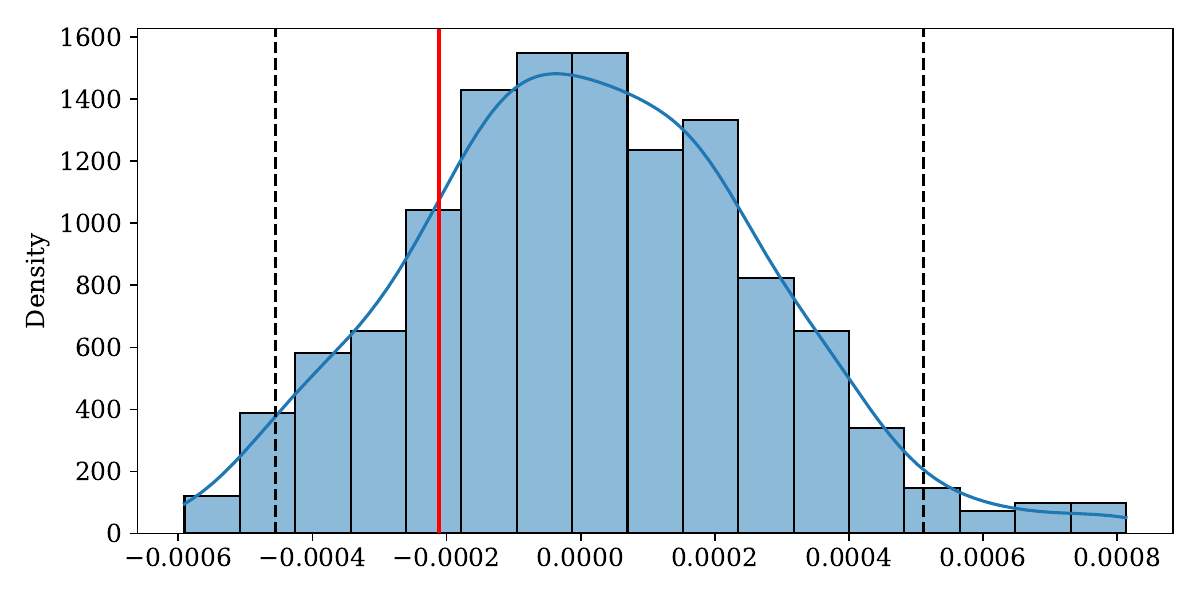}
\caption{Comparing $X_1$ vs. $X_3$}
    \label{fig:test1}
  \end{subfigure}%
  \hfill
  \begin{subfigure}{.5\linewidth}
    \centering
    \includegraphics[width=\linewidth]{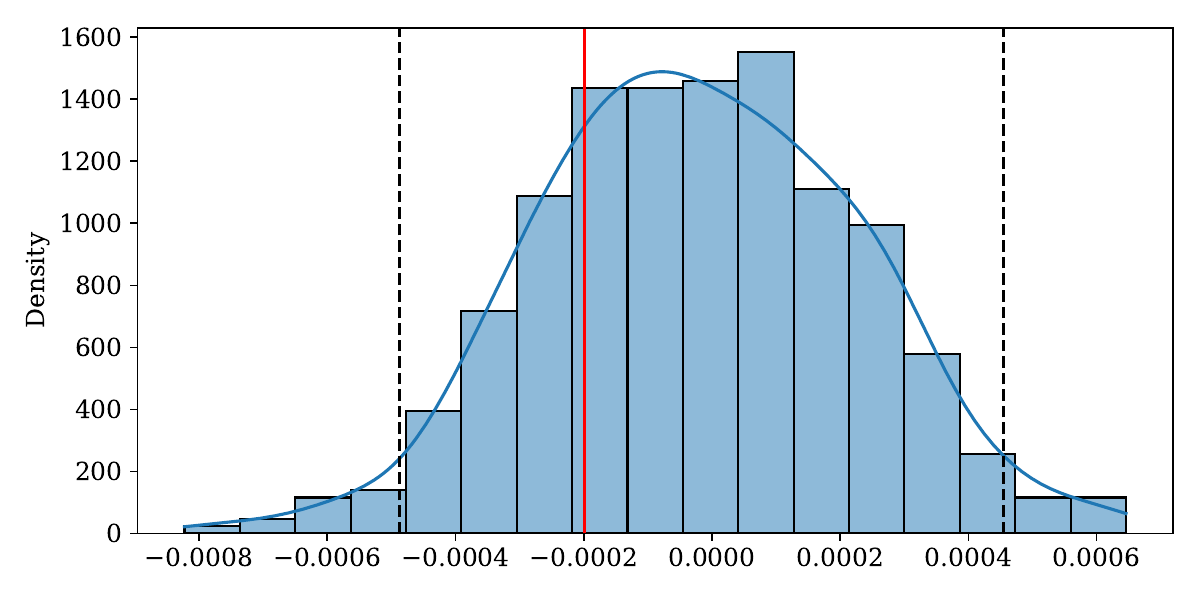}
\caption{Comparing $X_2$ vs. $X_4$}
    \label{fig:test2}
  \end{subfigure}
     \begin{subfigure}{.5\linewidth}
    \centering
    \includegraphics[width=\linewidth]{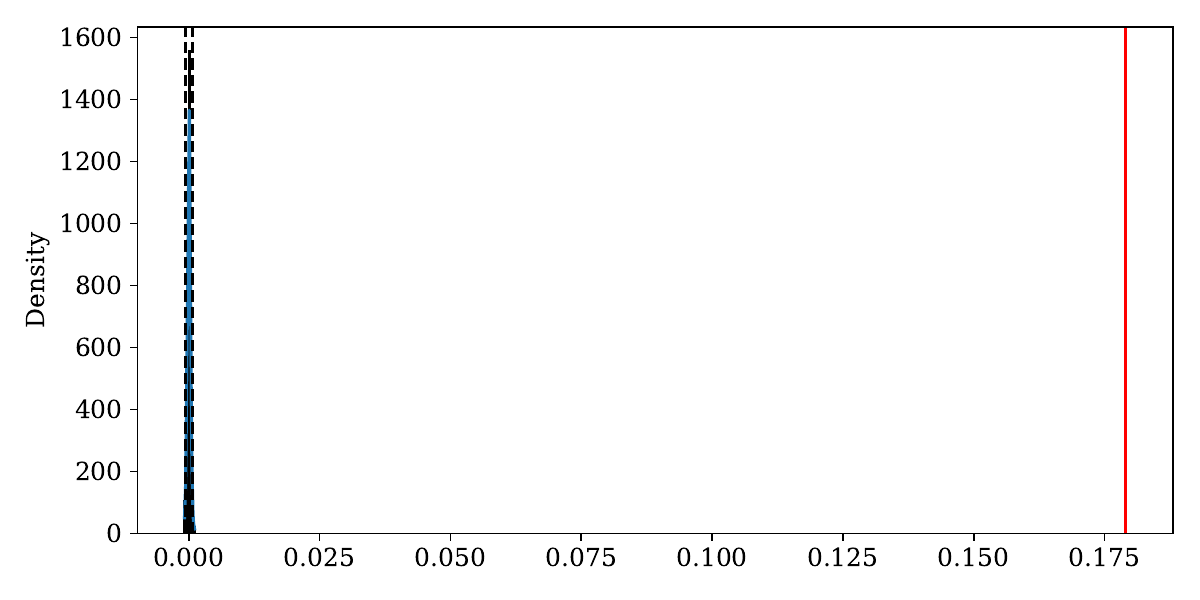}
\caption{Comparing $X_1$ vs. $X_2$ or $X_3$ vs. $X_4$}
    \label{fig:test3}
  \end{subfigure}%
  \vspace{-0.5em} 
  \caption{Hypothesis Testing: the null-distribution is in blue; the difference between the KIPC for the two reservoirs being compared ($\Delta$) is indicated by the red-continuous  vertical line; the $\alpha/2$ and $1-\alpha/2$ percentiles are indicated by the dashed black vertical lines. Notice that we cannot reject the null-hypothesis (with significance $\alpha$) for the comparisons $X_1,X_3$ and $X_2,X_4$. However, we can reject the null for comparisons $X_1,X_2$ and $X_3,X_4$.}
  \label{fig:test}
\end{figure*}
\end{example}

Therefore, we have also resolved \textbf{Issue 3}. By adopting the kernel approach, we perform a single joint hypothesis test to assess the significance of differences in capacity. Conversely, computing total IPC requires a separate hypothesis test for each term. Evaluating millions of such terms necessitates corrections for multiple comparisons, which can either severely impair statistical power if properly applied, or artificially inflate the total IPC if omitted.

\begin{remark}
\label{rem:2}
When using KIPC, it is often desirable to establish a threshold of significance  (that is, a capacity difference that can be statistically assessed) for a single isolated system (e.g., $\mathcal{S}_a$), without explicitly requiring a comparative baseline against a second system ($\mathcal{S}_a$ vs. $\mathcal{S}_b$). To compute such threshold of significance, in practice, we can evaluate a small number of samples $d_\pi$ of
\begin{align}
\label{eq:KIPCdelta2}
d_\pi &:= \frac{1}{n_T - 1} \sum_{i=1}^{n_T} \sum_{j=1,,j \neq i}^{n_T} \Pi^{(a)}_{ij} K{\otimes}(U_\pi, U_\pi)_{ij}
\end{align}
by using different random permutations, to estimate the order of magnitude required for a capacity difference to be statistically significant. This informal (more conservative) approach corresponds to evaluating $\Pi^{(a)}$ against an ideal baseline $\Pi^{(b)}$ with a smaller or equal rank.
\end{remark}

\subsection{The choice of $\rho$}
\label{sec:rho}
As discussed in Remark \ref{rem:1}, in KIPC, we make Assumption \ref{ass:1} explicit by choosing weights $\lambda_{\boldsymbol{\iota}}$ that decrease with the degree. For the Legendre kernel, this decay is given by $\rho^i/(2i+1)$, where $i$ denotes the degree of the polynomial. This weighting is explicit, in the sense that, for instance, the polynomial $\mathcal{L}_1(u_{t-1})\mathcal{L}_1(u_{t-2})\mathcal{L}_3(u_{t-1})$ has weight $\left(\frac{\rho}{3}\right)\left(\frac{\rho}{3}\right)\left(\frac{\rho^3}{7}\right)$.

The weights depend on a parameter $\rho\in(0,1)$. Ideally, one would choose $\rho\approx 1$, but when $ \rho\to 1$ this leads to numerical instability.
Note indeed that, in \eqref{eq:leg}, for $u_t=u_{t'}=1$ and $\rho=1$, $K_{Leg}(1,1)=\infty$.
For $\rho \approx 1$, this numerical instability produces an increase in the significance threshold in the hypothesis testing discussed in Section \ref{sec:hypo}, requiring a greater $n_T$ to assess significance. Therefore, to avoid this issue,  we set $\rho = 0.95$ in our experiments. This implies, for instance, that a nonlinearity of degree $50$ receives a weight of $\rho^{50}/101 \approx 0.000762$. To reliably distinguish reservoirs with differences on the order of $0.000762$, $n_T$ must be chosen so that these variations are statistically significant. We can determine a lower bound for $n_T$ by using the method described in Section \ref{sec:hypo}.
\begin{example}
We consider again \ref{ex:1} and compare
$$
\begin{aligned}
X_1 &= [\mathcal{L}_1(U_{\cdot 1}),\, \mathcal{L}_2(U_{\cdot 1}),\, \mathcal{L}_2(U_{\cdot 1})\mathcal{L}_1(U_{\cdot 2})],\\
X_5 &= [\mathcal{L}_1(U_{\cdot 1}),\, \mathcal{L}_2(U_{\cdot 1}),\, \mathcal{L}_2(U_{\cdot 1})\mathcal{L}_1(U_{\cdot 2}),\, \mathcal{L}_{50}(U_{\cdot 2})],
\end{aligned}
$$
The theoretical KIPC is: $ 0.554325$ (i=1) and $\approx 0.554325+0.000762=0.555087$ (i=5).
We select $n_T=10^4$ and perform a hypothesis testing with $\alpha=0.05$ and $n_{samples}=300$ to test the difference in capacity between these two reservoirs.  Figure \ref{fig:test2h1} shows that, for $n_T=10^4$, we can just reject the null-hypothesis (with significance $\alpha$) for the comparison $X_1,X_5$. The difference is clearly significant for $n_T=2.5\cdot 10^4$ as shown in Figure \ref{fig:test2h2} and certainly significant 
for $n_T=10^5$ as shown in Figure \ref{fig:test2h3}. 
We also estimate the significance based on the procedure described in Remark \ref{rem:2}, which only uses one reservoir matrix -- the one with higher-rank, $X_5$ in this case.
Using $30$ samples, the estimated significance at $\alpha=0.05$ for $|\Delta|$ is equal to $0.00103$ ($n_T=10^4$) and $0.00041$ ($n_T=5 \cdot 10^4$)
and $0.00008$ ($n_T=10^5$). With this more conservative approach, we can see that $n_T=10^5$ is more than enough to detect the nonlinearity $\rho^{50}/101 \approx 0.000762$.

\begin{figure*}[hbt!]
  \centering
   \begin{subfigure}{.5\linewidth}
    \centering
    \includegraphics[width=\linewidth]{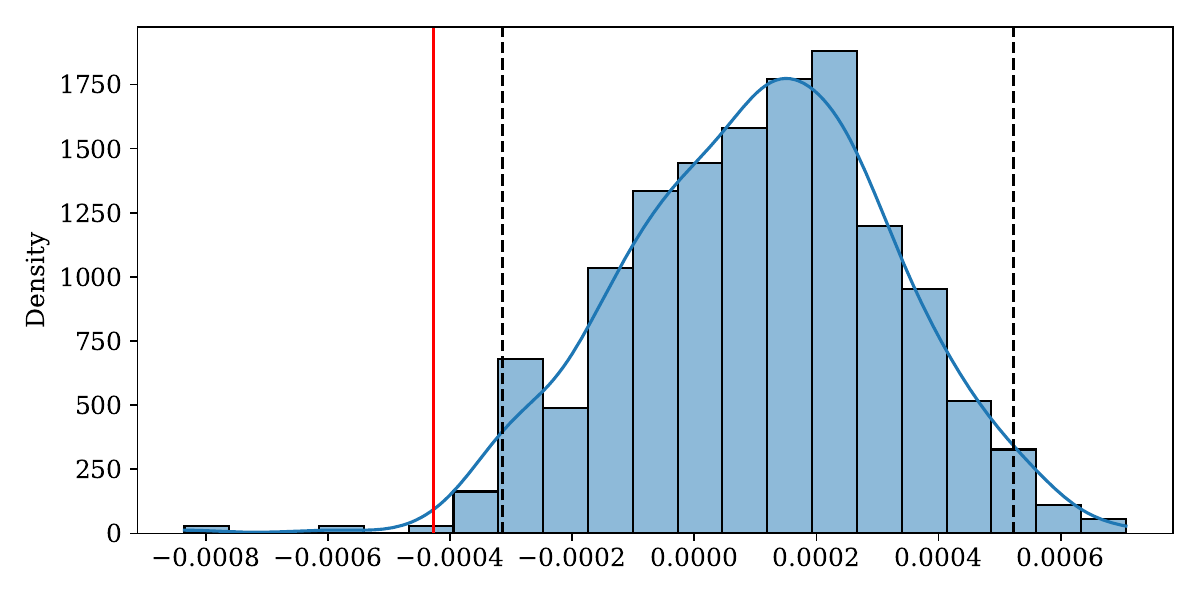}
\caption{$n_T=10^4$}
    \label{fig:test2h1}
  \end{subfigure}%
  \hfill
  \begin{subfigure}{.5\linewidth}
    \centering
    \includegraphics[width=\linewidth]{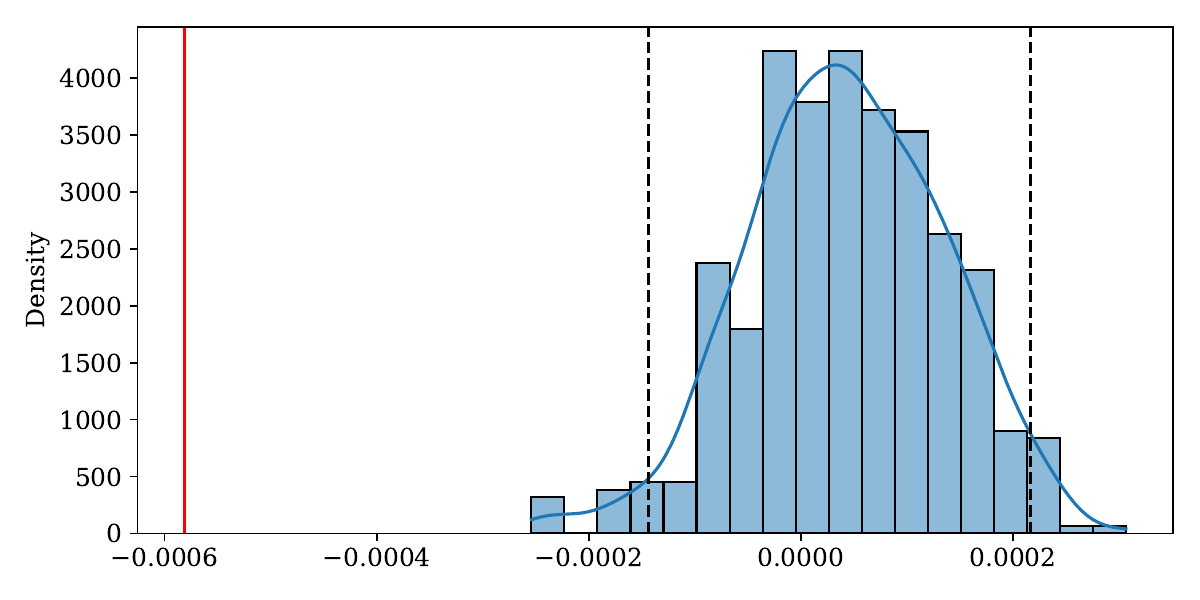}
\caption{$n_T=2.5\cdot 10^4$}
    \label{fig:test2h2}
  \end{subfigure}
      \begin{subfigure}{.5\linewidth}
    \centering
    \includegraphics[width=\linewidth]{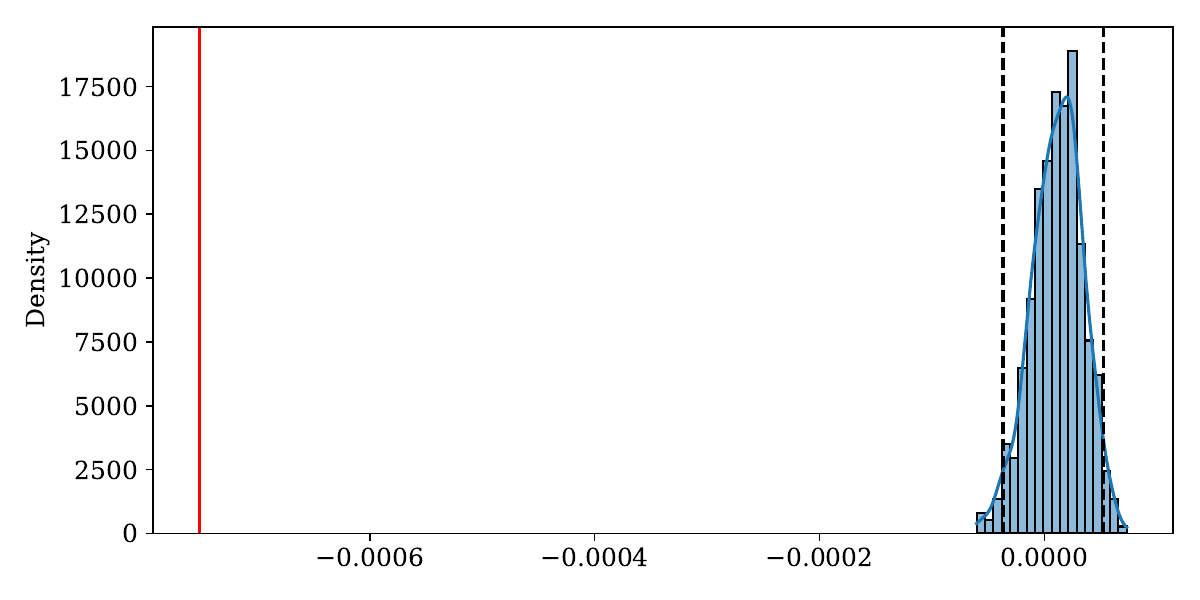}
\caption{$n_T=10^5$}
    \label{fig:test2h3}
  \end{subfigure} 
  \vspace{-0.5em} 
  \caption{Hypothesis Testing for $X_1$ vs.\ $X_5$: the null-distribution is in blue; the difference between the KIPC for the two reservoirs being compared ($\Delta$) is indicated by the red-continuous  vertical line; the $\alpha/2$ and $1-\alpha/2$ percentiles are indicated by the dashed black vertical lines. Notice that we can reject the null-hypothesis in each of the comparisons between $X_1$ and $X_5$. }
  \label{fig:test2h}
\end{figure*}

\end{example}

Assumption 1 also states that capacities should decrease with the delay. In KIPC, this can be achieved by using a different $\rho$ for each kernel in the product in \eqref{eq:kernelindprod}. For instance, considering again  the polynomial $\mathcal{L}_1(u_{t-1})\mathcal{L}_1(u_{t-2})\mathcal{L}_3(u_{t-1})$, we would have the weight $\left(\frac{\rho_1}{3}\right)\left(\frac{\rho_2}{3}\right)\left(\frac{\rho_1^3}{7}\right)$, with $\rho_1>\rho_2$, where $\rho_\tau$ is the weight corresponding to the delay $\tau$. While this is possible and straightforward, we do not implement it in this paper and instead use the same $\rho$ for each delay.

\subsection{Gram matrix}
For finite $n_T$, when evaluating \(\frac{1}{n_T}{Tr}(\Pi K_{\otimes}(U,U))\), we are in practice computing 
$\frac{1}{n_T}\operatorname{Tr}\bigl(\Pi \sum_{i=1}^{n_T} \gamma_i \mathbf{v}_i \mathbf{v}_i^\top \bigr)$, 
where $(\gamma_i, \mathbf{v}_i)$ denotes the eigenvalue-eigenvector decomposition of the kernel matrix (Gram matrix) $K_{\otimes}(U,U)$. 
Therefore, with KIPC we always evaluate the information capacity with respect to an orthogonal basis. 
This property holds even when the input signal $u$ is not distributed according to the reference measure that orthogonalises the underlying polynomials defining the kernel (e.g., $u \sim \operatorname{Unif}(-1,1)$ for the Legendre kernel). 
Consequently, KIPC automatically resolves \textbf{Issue 4}.

\section{Numerical experiments}
In the following numerical experiments, we will always use the Legendre kernel defined in~\eqref{eq:leg}. We evaluate the information processing capacity of a classical RC, an \textit{echo-state network}. The goal is to show the issues discussed for IPC and  how KIPC addresses them. This will demonstrate the greater reliability as well as the greater efficiency (in computational time) of KIPC. We will then use KIPC to quantify the information processing capacity of a QRC, specifically a \textit{disordered quantum spin chain reservoir}.

\subsection{Echo-state network}\label{sec:echo state network}
An echo-state network (ESN) is defined as \cite{jaeger2004harnessing}:
\begin{align}
{\bf x}_{t+1} &= \mathrm{tanh}\left(s W_2 {\bf x}_{t} + 0.7\tilde{\bf u}_{t+1}\right)\\
\tilde{\bf u}_{t+1} &= W_1 (0.5u_{t+1}+0.5)^p
\end{align}
where $u \in \text{Unif}(-1,1)$, $W_1 \in \mathbb{R}^{n_s \times 1}$, and $W_2 \in \mathbb{R}^{n_s \times n_s}$ have elements randomly drawn from $\text{Unif}(-1,1)$, $s>0$ is the spectral radius, and the number of states is $n_s=50$. We evaluate the IPC across $s \in \{0.1,0.2,\dots,1.7\}$ for $p=1$ (standard ESN) and $p=9$ (highly nonlinear ESN). Note that setting $p=9$ means applying a nonlinear (polynomial) transformation to the input signal before feeding it into the ESN. This is not standard practice in ESNs; we adopt it here because it allows us to clearly illustrate \textbf{Issues 1--3} that affect the computation of total IPC.

We run simulations for 210,000 time steps, discarding the first 10,000 as warm-up and repeat the experiment for $MC=10$ (Monte Carlo) runs.

For standard total-IPC, we consider all polynomial degrees from 1 to 10 with corresponding maximum delays set to
\begin{equation}
\label{eq:IPClist}
    \{30, 30, 30, 27, 25, 20, 17, 14, 11, 10\}.
\end{equation}
The reduction of the delay at the increase of the polynomial degree  is standard practice in the computation of total-IPC -- it is necessary to prevent the combinatorial explosion in the number of terms (\textbf{Issue 1}). For KIPC, we only fix the maximum delay to $30$, as it implicitly accounts for all degrees up to infinity.
The  total-IPC is upper bounded by 50 (the rank of the matrix $X$). For KIPC, the upper bound is 
13.11 from Theorem \ref{th:2} (discarding the contribution of the constant term, which we do not include, as discussed in Section~\ref{sec:ipc}, assuming the columns of the matrix $X$ have been centred to have zero mean).

Figure~\ref{fig:a1} summarises the results. The left column of Figure~\ref{fig:a1} shows the results for $p = 1$. For the total-IPC, shown in Figures~\ref{fig:a11} and~\ref{fig:a13}, we report the usual breakdown of IPC contributions by nonlinearity order -- we computed it using the Python \textit{ipc-module} \cite{inoue2026reservoir}.
Each color represents the IPC term $\mathrm{C}^d(\mathcal{S}, \mathcal{L}_d, u) $ relative to the degree $d=1,\dots,10$. They are plotted one above the other, so that their upper envelope corresponds to the total-IPC.

Figures~\ref{fig:a11} and~\ref{fig:a13} differ only in their threshold scaling factors (3 and the more aggressive 5, respectively), which are used to filter out spurious correlations in the high-degree terms, as described in \textbf{Issue 3}. The total-IPC (the upper envelope of these curves) can be compared to KIPC in Figure~\ref{fig:a15}, where it peaks for a spectral radius $s \in [0.9, 1]$ and then declines for $s > 1$ due to unstable ESN dynamics. 
Note that, also for KIPC we reported $KIPC_{lin}$, which corresponds to  $\mathrm{C}^1(\mathcal{S}, \mathcal{L}_1, u) $ in total-IPC.

Comparing Figure~\ref{fig:a11} with Figure~\ref{fig:a15}, it is evident that total-IPC underestimates the capacity when the spectral radius approaches the edge of instability ($s=1$).
Indeed, KIPC in Figure~\ref{fig:a15} indicates that the capacity reaches its maximum at $s=1$, whereas,  according to total-IPC, the capacity begins to decrease already before reaching $s=1$. As $s\to 1$, the system’s response to the input becomes increasingly nonlinear, and this behavior is not well captured by total-IPC due to \textbf{Issue 1} and \textbf{Issue 3}. This drawback for total-IPC was already observed by \citet{dambre2012information}. In contrast, KIPC can detect the increase of nonlinear behaviour as $s\to 1$.

For $p = 9$, a threshold scaling factor of $3$ is insufficient to eliminate the spurious correlations that unphysically push the total IPC above the ESN rank limit of 50 (Figure~\ref{fig:a12}).  We have to use the stronger threshold factor of $5$ to get rid of this unphysical behaviour, however in this way we also remove non-spurious IPC terms underestimating the IPC, \textbf{Issue 3}. This is evident from the comparison of Figure~\ref{fig:a12} and Figure~\ref{fig:a14}.
The KIPC in Figure~\ref{fig:a16} instead evaluates all degree terms rather than capping them at degree 10 and moreover the uncertainty (vertical bars) is very low in the stable dynamics ESN  region ($s < 1$) showing that KIPC is much more reliable than total-IPC.
Note that, for simplicity of notation, when we refer to KIPC and KIPC$_{\text{lin}}$ we mean $\widehat{\mathrm{KIPC}}$ and $\widehat{\mathrm{KIPC}}_{\text{lin}}$, so we omit the hat notation hereafter.

Comparing Figure~\ref{fig:a14} with Figure~\ref{fig:a16} for $s=1$, it is even more evident that the nonlinear behavior of ESN when it approaches $s=1$ is not well captured by total-IPC. Indeed the total-IPC for $s=1$ is 35, which is way below the expected maximum $50$. Instead, KIPC tells us that the maximum information capacity is still reached for $s=1$.

\begin{figure}[hbt!]
  \centering
    \includegraphics[width=\linewidth]{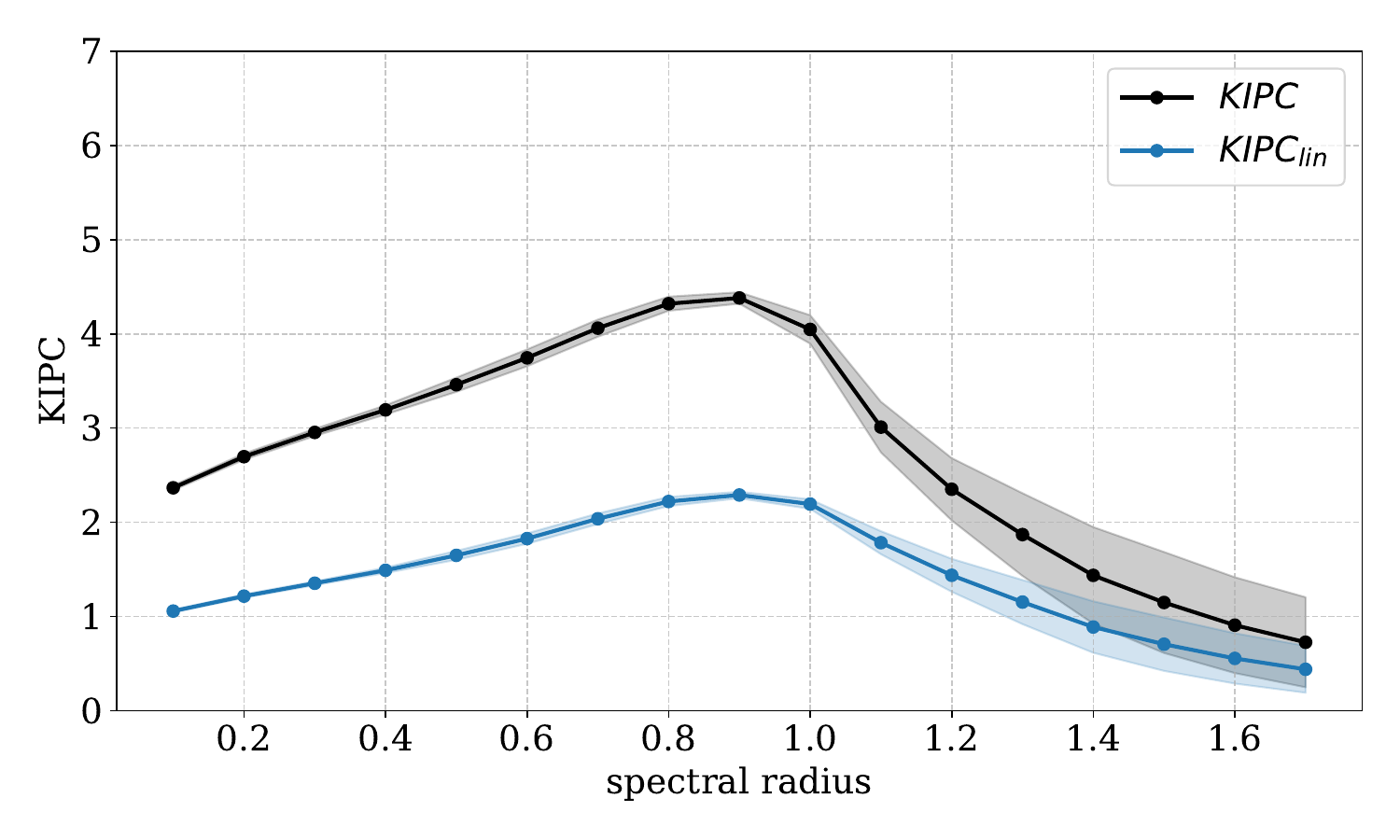}
\caption{$p=9$: KIPC with $\tau_{max}=15$. Notice that the capacity computed by KIPC starts to decrease at $s=0.9$. Since $KIPC_{\text{lin}}$ only marginally exhibits this decrease, the reduction of $\tau_{\max}$ primarily affects the nonlinear terms.}
 \label{fig:c3}
\end{figure}

\begin{figure*}[hbt!]
  \centering
  \begin{subfigure}{.5\linewidth}
    \centering
    \includegraphics[width=\linewidth]{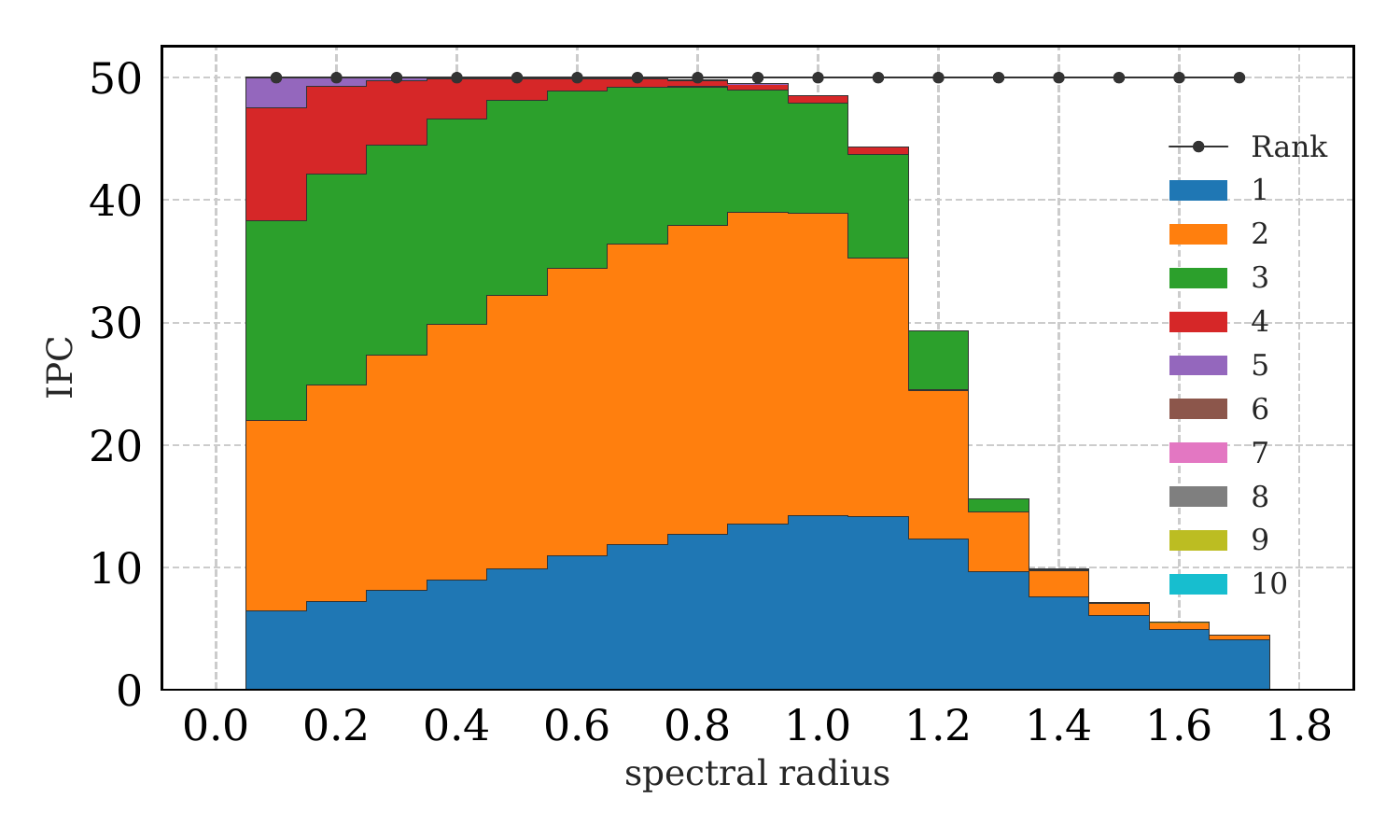}
\caption{$p=1$: IPC with threshold scaled by 3}
    \label{fig:a11}
  \end{subfigure}%
  \hfill
  \begin{subfigure}{.5\linewidth}
    \centering
    \includegraphics[width=\linewidth]{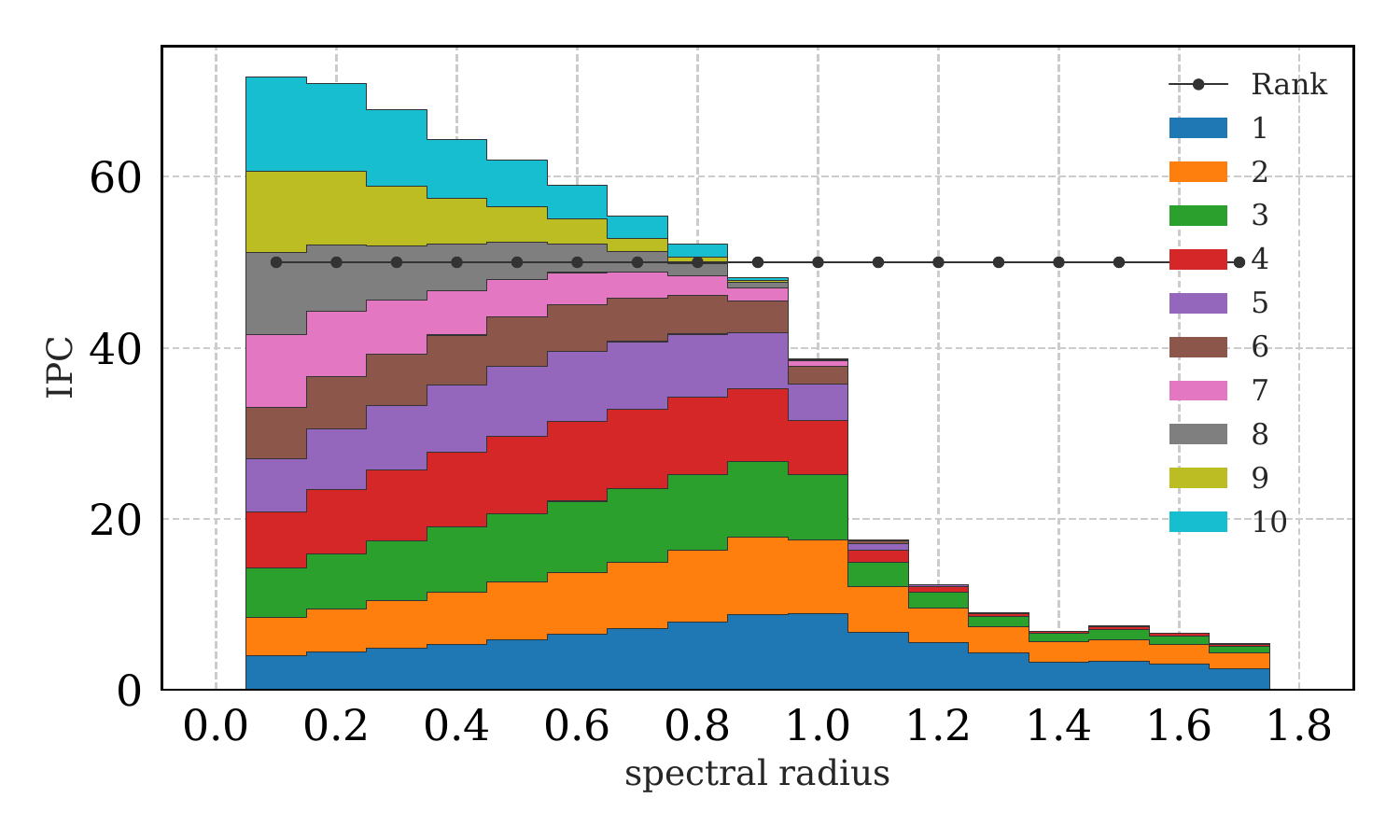}
\caption{$p=9$: IPC  with threshold scaled by  3}
    \label{fig:a12}
  \end{subfigure}
    \begin{subfigure}{.5\linewidth}
    \centering
    \includegraphics[width=\linewidth]{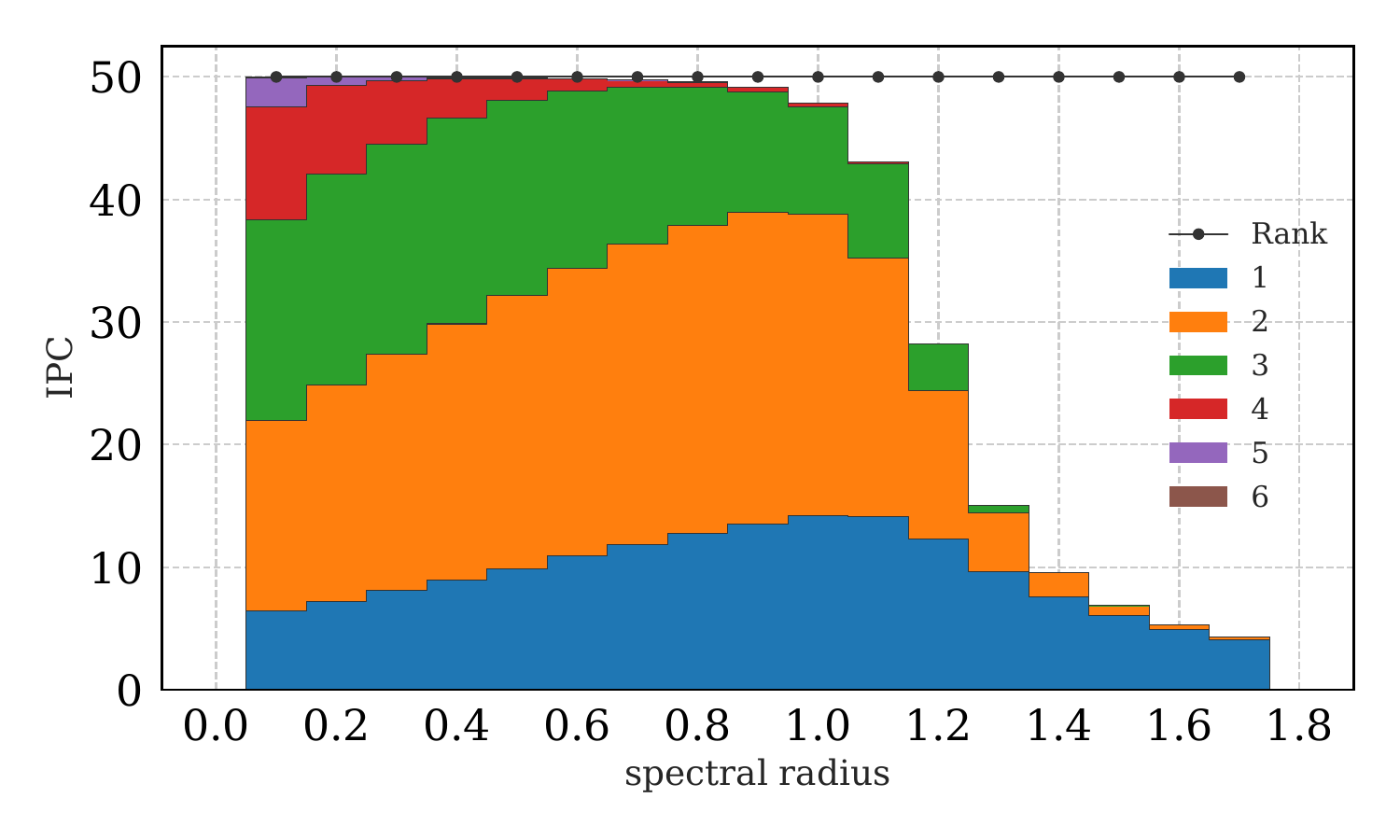}
\caption{$p=1$: IPC  with threshold scaled by 5}
    \label{fig:a13}
  \end{subfigure}%
  \hfill
  \begin{subfigure}{.5\linewidth}
    \centering
    \includegraphics[width=\linewidth]{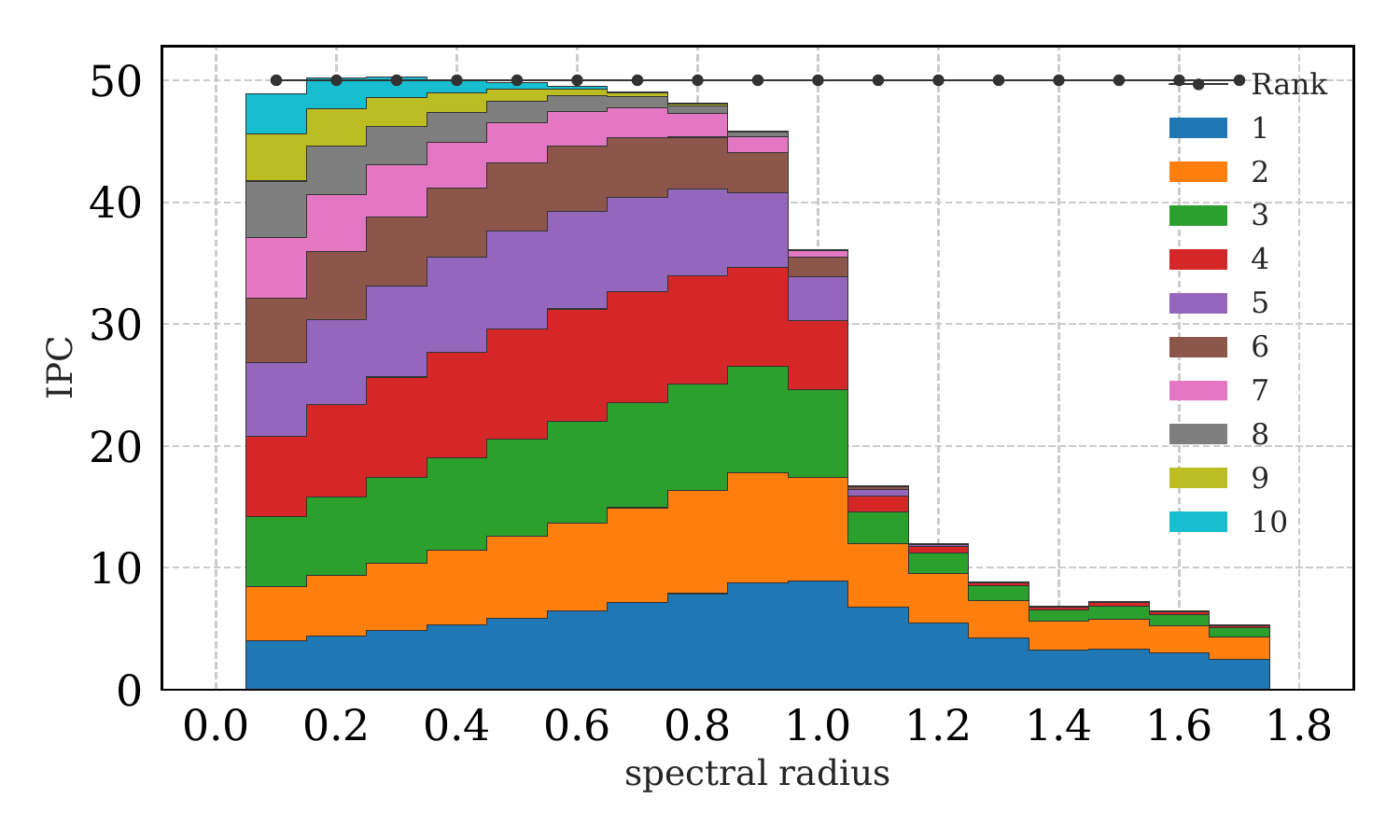}
\caption{$p=9$: IPC  with threshold scaled by 5}
    \label{fig:a14}
  \end{subfigure}

  \vspace{-0.5em} 

  \begin{subfigure}{.5\linewidth}
    \centering
    \includegraphics[width=\linewidth]{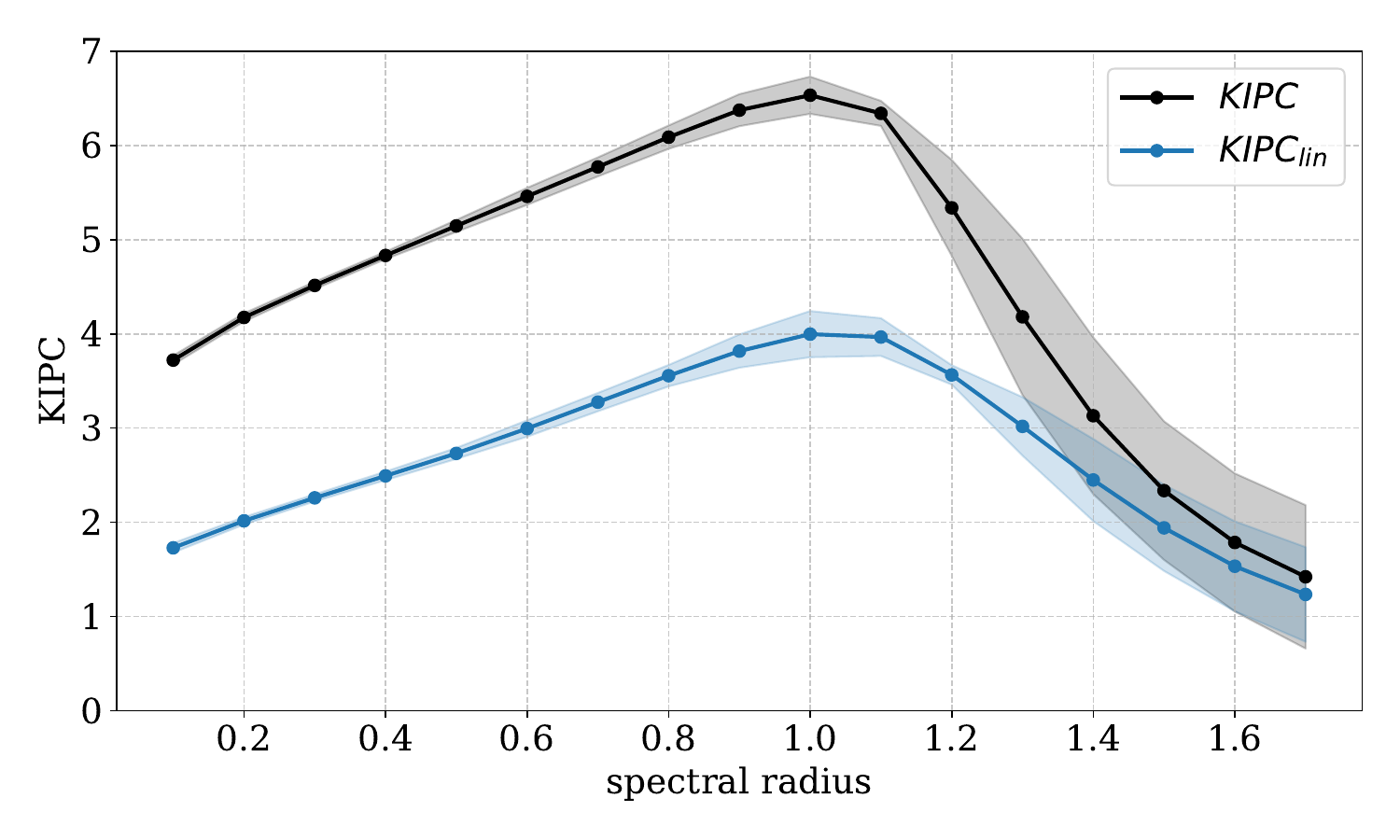}
\caption{$p=1$: KIPC}
    \label{fig:a15}
  \end{subfigure}%
  \hfill
  \begin{subfigure}{.5\linewidth}
    \centering
    \includegraphics[width=\linewidth]{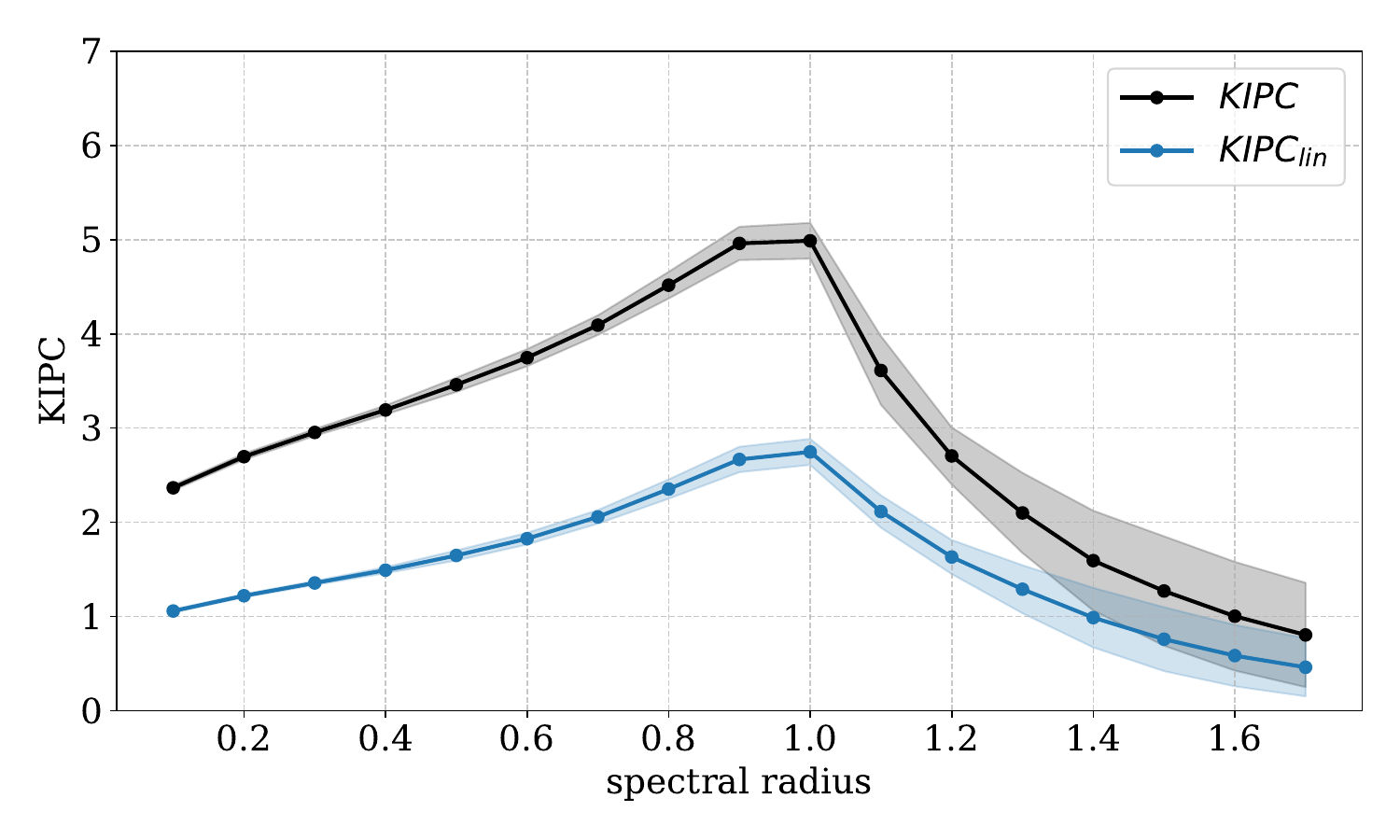}
\caption{$p=9$: KIPC}
    \label{fig:a16}
  \end{subfigure}

  \vspace{-0.5em} 
  \caption{Computation of total-IPC and KIPC for the ESN numerical experiment. The left-column plots are relative to the case $p=1$, while the right-column plots to $p=9$. Section \ref{sec:echo state network} contains a detailed analysis of the plots. In the KIPC-plots,  shaded regions represent 95\% confidence intervals computed over the Monte Carlo runs.}
  \label{fig:a1}
\end{figure*}

To validate that this behaviour of the total-IPC is due to the truncation (\textbf{Issue 1}), we re-run KIPC for $p=9$ with a maximum delay of $\tau_{\max}=15$, which makes the computed values more consistent with the truncation in \eqref{eq:IPClist} used to compute the total-IPC.
Figure \ref{fig:c3} shows that the capacity computed by KIPC starts to decrease at $s=0.9$. Since $KIPC_{\text{lin}}$ only marginally exhibits this decrease, the reduction of $\tau_{\max}$ primarily affects the nonlinear terms. This clearly explains why the total-IPC underestimates the capacity at $s=1$.

Comparing Figure~\ref{fig:a15} with Figure~\ref{fig:a16}, we observe that the KIPC values for $p=9$ are lower than those for $p=1$. This is fully consistent with Assumption~1, which favours systems with lower linearity over highly nonlinear ones (therefore, addressing \textbf{Issue 2}). Indeed, the standard implementation of an ESN does not apply a polynomial transformation to the inputs.

Finally, we report the computation time (on the same GPU as in Example \ref{ex:1}) required to compute the total IPC and KIPC in Figure \ref{fig:a1} for each value of the spectral radius (with $n_T = 200,000$). It can be noted that computing KIPC is 10-times faster -- it also yields correct results, since it does not truncate the maximum polynomial degree.
\begin{table}[ht]
    \centering
    \begin{tabular}{c|ccc}
        & Stand.\ IPC & KIPC-Leg \\
        \hline
        time (sec) & 2750 & 280 \\
    \end{tabular}
    \caption{Computational time comparison between standard IPC and KIPC-Leg. Notice that computing KIPC is 10-times faster. }
    \label{tab:computational-time}
\end{table}

\subsection{Disordered quantum spin chain reservoir}\label{sec:disordered quantum}
We consider the disordered all-to-all transverse-field Ising model as the substrate for the quantum reservoir (QRC) numerical application. This choice is motivated by its rich dynamics as the transverse-field strength $h$ and disorder amplitude $W$ are varied. Several works have shown that these regimes strongly influence QRC performance \cite{martinez2021dynamical,palacios2024role,keenan2026storage,yuan2022quantum}, using benchmarks (such as NARMA and delay) and standard total IPC \cite{keenan2026storage} (but restricting the polynomial degree to at most 4). Here, we aim to strengthen this analysis by using KIPC to shed additional light on the performance of this QRC.

The reservoir is defined by the Hamiltonian with on-site disorder \cite{martinez2021dynamical}:
\begin{equation}
H = \sum_{i>j=1}^{N} J_{ij}\,\sigma_i^{x}\sigma_j^{x} 
    + \frac{1}{2}\sum_{i=1}^{N} (h + w_i)\,\sigma_i^{z},
\label{eq:tfim_hamiltonian}
\end{equation}
where $N$ is the number of spins, $h$ is the magnetic field, $w_i$ is the on-site disorder, $\sigma_i^{a}$ ($a=x,y,z$) are the Pauli matrices, and $J_{ij}$ are the spin--spin couplings. To generate the reservoir, the parameters are randomly generated $J_{ij} \in \text{Unif}(-1/2,1/2)$  and $w_i \in \text{Unif}(-W,W)$.
Following the approach of previous work, we consider two parameter sweeps. In case 1, we vary the average transverse-field strength $h$ at vanishing disorder ($W=0$). In case 2, we vary the disorder strength $W$ at fixed average field ($h=10$). Phase-diagram analyses indicate \cite{martinez2021dynamical} that, across these sweeps, the system undergoes transitions between distinct dynamical phases. We therefore expect the IPC to vary over the sweeps.

We recall that a QRC algorithm can be divided into three steps: 
(i) inject an input into the reservoir system; 
(ii) let the reservoir (i.e., the spin network defined in  \eqref{eq:tfim_hamiltonian}) evolve; (iii) extract information from the reservoir using all or some of its degrees of freedom via an output layer.

We inject \cite{martinez2021dynamical} the input signal  $\{u_1, u_2, \dots, u_{n_T}\}$, with $u_k \in \text{Unif}(-1,1)$, into the qubit~1 at every time step $k$. This spin state is updated every $\Delta t=10$ steps as
\[
\rho^{(k)}_1 = |\psi_{\tilde{u}_k}\rangle\langle\psi_{\tilde{u}_k}|,
\quad
|\psi_{\tilde{u}_k}\rangle := \sqrt{1 - \tilde{u}_k}\,|0\rangle + \sqrt{\tilde{u}_k}\,|1\rangle,
\]
where $\tilde{u}=0.5(u_k+1) \in [0,1]$.
The reservoir is then let to evolve:
\begin{equation}
\rho(k\Delta t) 
= e^{-i H \Delta t}
\Bigl( \rho^{(k)}_1 \otimes \operatorname{Tr}_1\big(\rho((k-1)\Delta t)\big) \Bigr)
e^{i H \Delta t},
\label{eq:cptp_map}
\end{equation}
where $e^{-i H \Delta t}$ is the unitary dynamics operator and $\operatorname{Tr}_1\{\cdot\}$ denotes the partial trace over the first qubit. 

For the output layer we use the following observables \cite{martinez2021dynamical}: the single-qubit expectations $\langle \sigma^x_i \rangle$, $\langle \sigma^y_i \rangle$, and $\langle \sigma^z_i \rangle$ for each qubit $i$, together with all distinct two-qubit correlations $\langle \sigma^x_i \sigma^x_j \rangle$ with $i<j$. In practice, at each time step $k$ we first update $\rho(k\Delta t)$ via Eq.~\eqref{eq:cptp_map}, then compute the feature vector by evaluating $\langle O \rangle_k = \operatorname{Tr}(O\,\rho(k\Delta t))$ for each observable $O$; this corresponds to an idealised readout in which measurement back-action is neglected, whereas in a physical implementation the same statistics would be estimated by repeating the experiment over many runs and performing measurements on identically prepared copies of the reservoir.

In the simulations, we set $N=8$ qubits, yielding a total of 52 output signals. Consequently, the matrix $X$ in Theorem \ref{th:1} has rank 52, which make it comparable to the ESN considered in the previous section (rank 50).

We run simulations for 21,000 time steps, discarding the first 1,000 as a warm-up, and repeat the experiment for $MC=30$ Monte Carlo runs. Note that simulating the reservoir is computationally demanding due to the dimension of the density-matrix $\rho(k \delta t)$ of $2^8 \times 2^8$. For this reason, we limit $n_T$ to 20,000 when computing KIPC.

\begin{figure*}[hbt!]
  \centering
   \begin{subfigure}{.5\linewidth}
    \centering
    \includegraphics[width=\linewidth]{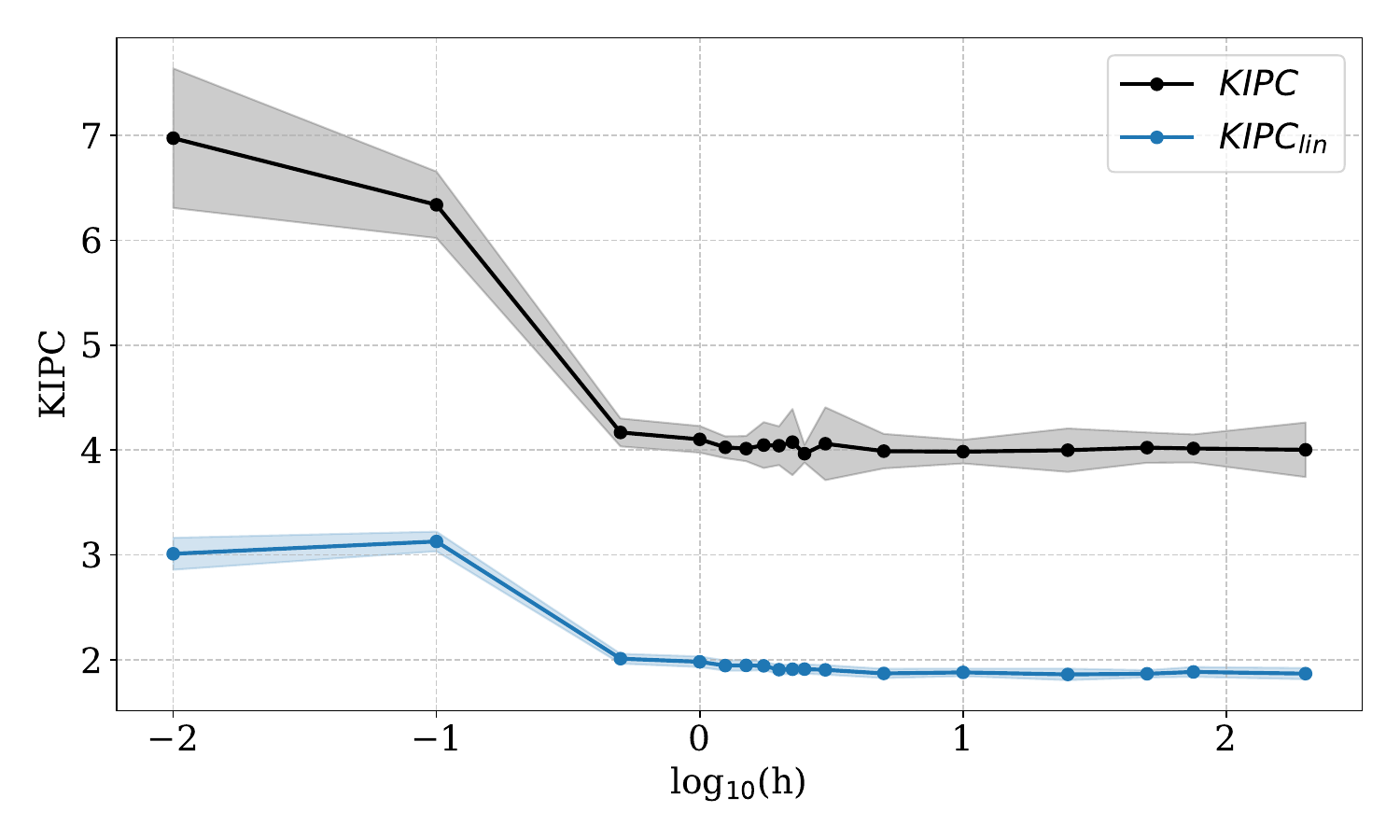}
\caption{case 1}
    \label{fig:b11}
  \end{subfigure}%
  \hfill
  \begin{subfigure}{.5\linewidth}
    \centering
    \includegraphics[width=\linewidth]{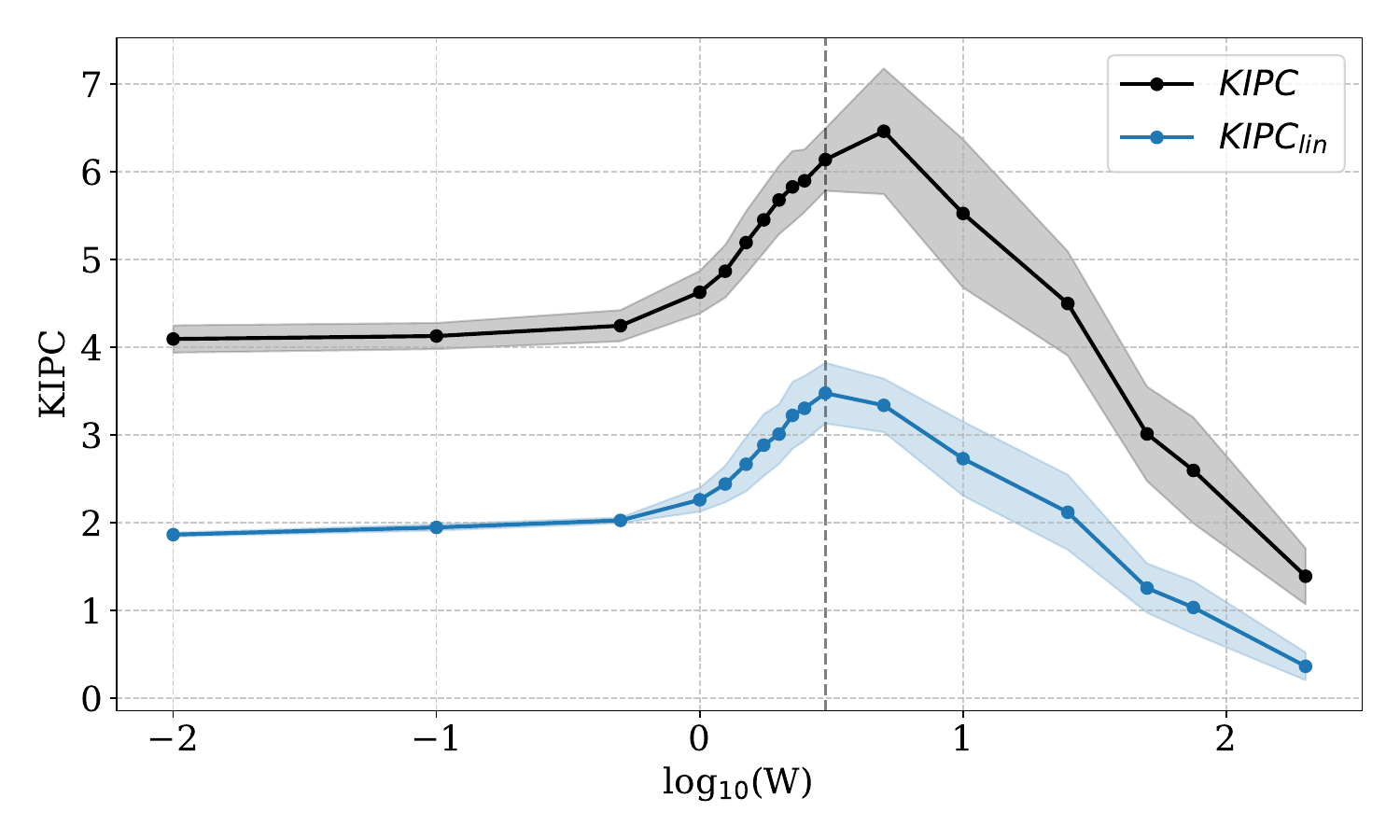}
\caption{case 2}
    \label{fig:b22}
  \end{subfigure}

  \vspace{-0.5em} 
  \caption{Computation of KIPC for the QRC numerical experiment. Shaded regions represent 95\% confidence intervals computed over the Monte Carlo runs. In subfigure \ref{fig:b11}, We observe that the KIPC (and KIPC$_{\mathrm{lin}}$) is large for small $h$ and then decreases for $h>0.1$ (phase transition), eventually reaching a plateau. In subfigure \ref{fig:b22}, we observe that the KIPC (and  KIPC$_{\mathrm{lin}}$) increases up to $W\in[3,5]$ (phase transition) and then decreases towards $0$. Section \ref{sec:disordered quantum} contains detailed analysis.}
  \label{fig:b1}
\end{figure*}

Figure \ref{fig:b1} reports the computed KIPC for the two cases

\begin{description}
    \item[case 1] We observe that the KIPC (and KIPC$_{\mathrm{lin}}$) is large for small $h$ and then decreases for $h>0.1$ (phase transition), eventually reaching a plateau. This behaviour is consistent with what is reported in \cite[Fig.~4]{martinez2021dynamical} for the squared Pearson correlation coefficient between the true and predicted output in the NARMA10 task.
    \item[case 2] We observe that the KIPC (and  KIPC$_{\mathrm{lin}}$) increases up to $W\in[3,5]$ (phase transition) and then decreases towards $0$. This trend also agrees with the NARMA10 results in \cite[Fig.~4]{martinez2021dynamical}. A key additional insight provided by KIPC is that, as $W$ increases, the nonlinear contribution
   $        \mathrm{KIPC}_{\mathrm{nonlin}} = \mathrm{KIPC} - \mathrm{KIPC}_{\mathrm{lin}}$
    grows up to $W=3$ (vertical line). For $W>3$, KIPC$_{\mathrm{lin}}$ starts to decrease while KIPC$_{\mathrm{nonlin}}$ continues to increase, leading to a maximum KIPC at $W=5$ (note that the exact maximiser may not be at $W=5$ due to the step size used in our experiments). This behaviour is interesting because, unlike in ESNs where KIPC and KIPC$_{\mathrm{lin}}$  peak at the same spectral radius, here the maxima do not coincide. To the best of our knowledge, this property of the `disordered quantum spin chain reservoir' has not been reported before. Overall, this illustrates that KIPC can provide new insights into QRC.
\end{description}

\subsection{Software}
The KIPC algorithm is implemented in Python. The source code for computing the metric and replicating the experiments is available at \url{https://github.com/benavoli/KernelIPC}.

\section{Conclusions}
In this work, we have introduced a kernel-based approach, the kernel-based information processing capacity (KIPC) to measure the information processing capacity metric of a dynamical system. Compared to the total-IPC metric introduced by \citet{dambre2012information}, our approach resolves several key limitations. First, we can compute the information capacity without imposing an upper bound on the polynomial degree, therefore accounting for all possible non-linearities. Moreover, this computation can be performed efficiently. Through a permutation-based test, we also show that we can statistically compare the KIPC of two dynamical systems using a single null-hypothesis test. This solves another prominent issue with the standard IPC, which instead requires a separate test for each element of the basis function. We demonstrated these improvements in practice using an Echo State Network as a test-bed dynamical system. Finally, we showed how the efficiency and reliability of KIPC can be leveraged to drive research in quantum reservoir computing (QRC), focusing specifically on a disordered quantum spin chain reservoir. While KIPC confirmed several previously known properties regarding the IPC for this QRC architecture, it also allowed us to discover an interesting phenomenon: a misalignment between the maximum of the linear KIPC component and the nonlinear KIPC components. This finding warrants further investigation.

As future work, we plan to replace the permutation testing we use for determining the distribution of the null-hypothesis with a parametric approximation of the null distribution computed analytically from its moments and, therefore, requiring only a small computational overhead compared to KIPC  (permutation testing instead requires computing the KIPC \(n_{\text{samples}}\) times). This is similar to the approach used in kernel-based independence testing, where a Gamma distribution is employed to approximate the null distribution \cite{gretton2007kernel}. Another valuable improvement would be to replace the null-hypothesis significance test with a Bayesian framework to more easily distinguish between statistical and practical significance, similar to the approach in \cite{9564124}. Finally, we plan to exploit the connection between IPC and Volterra series to investigate whether it is possible to tailor KIPC more closely to the computational characteristics of QRC. The work of \cite{benavoli2026quantum} provides a first step in this direction by offering a kernel interpretation for QRC architectures based on quantum linear dynamic networks.


\begin{acknowledgments}
We wish to acknowledge the support of the author community in using
REV\TeX{}, offering suggestions and encouragement, testing new versions,
\dots.
\end{acknowledgments}

\newpage
\appendix

\section{Appendixes}

\subsection{Proofs}
\label{app:proofs}
\paragraph{Proof of Corollary \ref{co:1}}
We  exploit the following integral representation 
$$
F_4\left(\tfrac{1}{2},1;1,1;X,Y\right)=\tfrac{1}{\pi}\int_0^\pi \frac{d\psi}{\sqrt{1-X-Y-2\sqrt{XY}\cos\psi}}
$$
which can be derived from \cite{schlosser2013multiple}. Since
\begin{align}
\nonumber
\tfrac{2}{\sqrt{p+q}}\,R\left(\sqrt{\tfrac{2q}{p+q}}\right)&=\int_{0}^{\pi} \tfrac{d \theta}{\sqrt{p+q-2q \sin^2(\theta/2)}}\\
\nonumber
&=\int_{0}^{\pi} \tfrac{d \theta}{\sqrt{p+q-q (\cos(\theta)-1)}}\\
&=\int_{0}^{\pi} \tfrac{d \theta}{\sqrt{p-q \cos(\theta)}},
\end{align}
we can then rewrite
\begin{align}
\nonumber
F_4\left(\tfrac{1}{2},1;1,1;X,Y\right)&=\tfrac{2}{\pi\sqrt{1-X-Y+2\sqrt{XY}}}\\
&R\left(\sqrt{\tfrac{4 XY}{1-X-Y+2\sqrt{XY}}}\right).
\end{align}

Since $X=\tfrac{(1-x)(1-x')r}{(1+r)^2}$, $Y=\tfrac{(1+x)(1+x')r}{(1+r)^2}$

\begin{align}
\nonumber
&F_4\left(\tfrac{1}{2},1;1,1;X,Y\right)\\
&=\tfrac{2(r+1)}{\pi\sqrt{1+r^2-2r xx'+2r\sqrt{(1-x^2)(1-{x'}^2)}}}\\
\nonumber
&\cdot R\left(\sqrt{\tfrac{4r(1-x^2)(1-{x'}^2)}{1+r^2-2r xx'+2r\sqrt{(1-x^2)(1-{x'}^2)}}}\right).
\end{align}
where $r=\rho$. By exploiting \eqref{eq:jac}, we can then simplify $(r+1)$ and prove the result \eqref{eq:leg}.\\

\paragraph{Proof of Theorem \ref{th:1}}
We  introduce the notation $X=[{\bf x}_1,\dots,{\bf x}_{n_T}]^\top \in \mathbb{R}^{n_T \times n_s}$  and $U=[{\bf u}_1,\dots,{\bf u}_{n_T}]^\top \in \mathbb{R}^{n_T \times \tau_{max}}$.
From \eqref{eq:KproductP}, we have that $K_{\otimes}(U,U)$ is a matrix whose $ij$-element is
$$
\{K_{\otimes}(U,U)\}_{ij}=K_{\otimes}({\bf u}_i,{\bf u}_j).
$$
 Its i-th eigenvalue and eigenvector are:
$$
K_{\otimes}(U,U) {\bf v}_i = \hat{\lambda}_i {\bf v}_i, 
$$
where ${\bf v}^\top_i {\bf v}_i=1$. It can then be shown, see for instance \cite[Sec. 4.3.2]{williams2006gaussian}, that, for finite $n_T$,
$$
\phi_i({\bf u}_j)\approx \sqrt{n_T}({\bf v}_i)_j,
$$
where $\phi_i$ is the i-th eigenfunction of the kernel. Since  ${\bf v}^\top_i K_{\otimes}(U,U){\bf v}_i = \hat{\lambda}_i=\frac{1}{n_T}\phi_i^\top(U) K_{\otimes}(U,U) \phi_i(U)  \approx \tfrac{\lambda_i}{n_T}$, where
$\lambda_i$ is the corresponding eigenvalue of the the kernel function. Therefore, for $n_t \to \infty$, $\frac{\hat{\lambda}_i}{n_T}$ is an estimator for the kernel function eigenvalue $\lambda_i$. 
Therefore, since
$$
Tr(\Pi K_{\otimes}(U,U))=\sum_{i=1}^{n_T} \hat{\lambda}_i {\bf v}^\top_i \Pi {\bf v}_i=\sum_{i=1}^{n_T} \hat{\lambda}_i \frac{{\bf v}^\top_i \Pi {\bf v}_i}{{\bf v}^\top_i {\bf v}_i},
$$
and, as shown in \eqref{eq:KproductP}, since  $\phi_{\iota}({\bf u}_j)$ is equal to $\overline{\mathcal{P}}^{\otimes}_{\boldsymbol{\iota}}({\bf u}_j)={\mathcal{P}}^{\otimes}_{\boldsymbol{\iota}}({\bf u}_j)/\sqrt{(\mathcal{P}^{\otimes}_{\boldsymbol{\iota}}({\bf u}_j))^\top \mathcal{P}^{\otimes}_{\boldsymbol{\iota}}({\bf u}_j)}$, we have that
\begin{equation}
\label{eq:KTPCproof}
\begin{aligned}
\lim_{n_T \rightarrow \infty} \frac{1}{n_T}Tr(\Pi K_{\otimes}(U,U))  &= \sum_{\boldsymbol{\iota} \in \mathbb{N}^m}  {\lambda}_{\boldsymbol{\iota}}\frac{(\mathcal{P}^{\otimes}_{\boldsymbol{\iota}}(U))^\top \Pi \mathcal{P}^{\otimes}_{\boldsymbol{\iota}}(U)}{(\mathcal{P}^{\otimes}_{\boldsymbol{\iota}}(U))^\top \mathcal{P}^{\otimes}_{\boldsymbol{\iota}}(U)}\\
\end{aligned}
\end{equation}
Now consider the IPC term:
\begin{equation}
\label{eq:IPCtermproof}
\begin{aligned}
 \mathrm{C}^d(\mathcal{S}, \mathcal{P}^{\otimes}_{\boldsymbol{\iota}}, u)= 1-\min_{\hat{\bf w},\hat{w}_0} \frac{\sum_{t=1}^T (\mathcal{P}^{\otimes}_{\boldsymbol{\iota}}({\bf u}_t)- \hat{\bf w}^\top {\bf x}_t - \hat{w}_0)^2}{\mathcal{P}^{\otimes}_{\boldsymbol{\iota}}(U)^\top \mathcal{P}^{\otimes}_{\boldsymbol{\iota}}(U)},
\end{aligned}
\end{equation}
From the properties of linear regression, we can write the optimal prediction
$$
Y =\hat{\bf w}^\top X+ \hat{w}_0= \Pi \mathcal{P}^{\otimes}_{\boldsymbol{\iota}}(U),
$$
where $\Pi=X (X^\top X)^{-1} X^\top$ is the projection matrix (optionally including a column of ones to account for the intercept term) and, therefore, 
\begin{align}
\nonumber
 &\mathrm{C}^d(\mathcal{S}, \mathcal{P}^{\otimes}_{\boldsymbol{\iota}}, u)= 1-\min_{\hat{\bf w},\hat{w}_0} \frac{\sum_{t=1}^T (\mathcal{P}^{\otimes}_{\boldsymbol{\iota}}({\bf u}_t)- \hat{\bf w}^\top {\bf x}_t - \hat{w}_0)^2}{\mathcal{P}^{\otimes}_{\boldsymbol{\iota}}(U)^\top \mathcal{P}^{\otimes}_{\boldsymbol{\iota}}(U)}\\
 \nonumber
 &=1-\frac{(\mathcal{P}^{\otimes}_{\boldsymbol{\iota}}(U)-\Pi \mathcal{P}^{\otimes}_{\boldsymbol{\iota}}(U))^\top (\mathcal{P}^{\otimes}_{\boldsymbol{\iota}}(U)-\Pi \mathcal{P}^{\otimes}_{\boldsymbol{\iota}}(U))}{\mathcal{P}^{\otimes}_{\boldsymbol{\iota}}(U)^\top \mathcal{P}^{\otimes}_{\boldsymbol{\iota}}(U)}\\
 \label{eq:IPCtermproof}
  &=\frac{\mathcal{P}^{\otimes}_{\boldsymbol{\iota}}(U)^\top \Pi \mathcal{P}^{\otimes}_{\boldsymbol{\iota}}(U)}{\mathcal{P}^{\otimes}_{\boldsymbol{\iota}}(U)^\top \mathcal{P}^{\otimes}_{\boldsymbol{\iota}}(U)},
\end{align}
where we have exploited $\Pi^2=\Pi$. From this result, it follows that
\begin{equation}
\label{eq:IPCtermproof1}
\begin{aligned}
 \lim_{n_T \rightarrow \infty} \frac{1}{n_T}Tr(\Pi K_{\otimes}(U,U))  &= \sum_{\boldsymbol{\iota} \in \mathbb{N}^m}  {\lambda}_{\boldsymbol{\iota}}\frac{(\mathcal{P}^{\otimes}_{\boldsymbol{\iota}}(U))^\top \Pi \mathcal{P}^{\otimes}_{\boldsymbol{\iota}}(U)}{(\mathcal{P}^{\otimes}_{\boldsymbol{\iota}}(U))^\top \mathcal{P}^{\otimes}_{\boldsymbol{\iota}}(U)}\\
 &=\sum_{\boldsymbol{\iota} \in \mathbb{N}^m}  {\lambda}_{\boldsymbol{\iota}} IPC_{\boldsymbol{\iota}}.
\end{aligned}
\end{equation}

\paragraph{Proof of Theorem \ref{th:2}}
The bound
\begin{equation}
\label{eq:asymlimproof}
\lim\limits_{n_T \rightarrow \infty} \frac{1}{n_T}Tr(\Pi K_{\otimes}(U,U))  \leq \sum_{n=0}^{n_s} {\lambda}_{n}.
\end{equation}
follows from Theorem \ref{th:1} by noticing that 
the left hand side is maximised by taking as projection matrix $\Pi$ the $n_s$ eigenvectors of $K_{\otimes}(U,U)$ corresponding to the largest eigenvalues.

\paragraph{Proof of Proposition \ref{prop:U}}
The proof follows directly from the decomposition as V-statistic:
\begin{align}
\nonumber
\frac{1}{n_T}\operatorname{Tr}(\Pi K_{\otimes}(U,U))&=\frac{1}{n_T^2}\sum_{i=1}^{n_T} n_T\Pi_{ii} K_{\otimes}(U,U)_{ii}\\
\nonumber
&+\frac{1}{n_T^2}\sum_{i=1}^{n_T} \sum_{j=1,\,j \neq i}^{n_T} n_T \Pi_{ij} K_{\otimes}(U,U)_{ij},
\end{align}
by noticing that the first term vanishes asymptotically and, therefore, introduces a bias for finite $n_T$. By neglecting the diagonal term, we obtain an unbiased estimator, a U-statistic:
\begin{align}
\nonumber
&\frac{1}{n_T}\operatorname{Tr}(\Pi K_{\otimes}(U,U))\\
\nonumber
&=\frac{1}{n_T(n_T-1)}\sum_{i=1}^{n_T} \sum_{j=1,\,j \neq i}^{n_T} n_T \Pi_{ij} K_{\otimes}(U,U)_{ij},
\end{align}
The non diagonal term includes $n_T(n_T-1)$ terms, this is the reason why we have changed the denominator in the first fraction.

\paragraph{Proof of Corollary \ref{co:2}}
It follows directly from the definition of $K^{(lin)}$ and Proposition \ref{prop:U}.

\subsection{Computation of KIPC}
\label{app:alg}
We can efficiently compute KIPC by working with $B \times B$ blocks of the kernel matrix $ K_{\otimes}(U,U)$. Additionally we use the SVD decomposition of $X=V \Sigma Z^\top$ to write $\Pi$ as $V V^\top$, so that
$Tr(\Pi K_{\otimes}(U,U))=Tr(V^\top K_{\otimes}(U,U) V)$, where the matrix in the second $Tr$ has dimensions $r \times r$, where $r \leq \tau_{max}$ is the rank of $X$. In the below algorithm, we assume that $n=kB$ for some integer $k$.
~\newline

\begin{algorithm}[H]
\caption{Block-wise Computation of the unbiased KIPC $\frac{1}{n_T-1}\operatorname{Tr}(\Pi K_{\otimes}(U,U))$}
\KwIn{Signal $u \in \mathbb{R}^{n}$, feature matrix $X \in \mathbb{R}^{n \times d}$, kernel evaluation function $K_\otimes(\cdot, \cdot)$, max delay $\tau_{\max}$, block size $B$}
\KwOut{Unbiased KIPC}

\BlankLine
\tcc{SVD decomposition so that $\Pi=VV^\top$}
$X_{\text{centered}} = X - \frac{1}{n} \mathbf{1}_n \mathbf{1}_n^\top X$\;
$(V,r) \gets \text{ComputeSVD}(X_{\text{centered}})$\;
\BlankLine
\tcc{Build delay embedding representation}
$U \gets \text{BuildDelayed}(u, \tau_{\max})$ \tcp*{$U \in \mathbb{R}^{n \times r}$}
$n, r \gets \text{Shape}(V)$\;
\tcc{Init. $V^\top K V$ accumulator matrix}
$A \gets \mathbf{0}_{r \times r}$ 

\BlankLine
\tcc{Block-wise processing of kernel blocks}
\For{$i \gets 0$ \KwTo $n-1$ {step} $B$}{
    $I_i \gets \{i, \dots, \min(i+B-1, n-1)\}$\;
    $X_i \gets V[I_i, :]$, \quad $Z_i \gets U[I_i, :]$\;
    $Y_i \gets \mathbf{0}_{|I_i| \times r}$\;
        \For{$j \gets 0$ \KwTo $n-1$ {step} $B$}{
        $I_j \gets \{j, \dots, \min(j+B-1, n-1)\}$\;
        $X_j \gets V[I_j, :]$, \quad $Z_j \gets U[I_j, :]$\;
                \BlankLine
        \tcc{Evaluate kernel matrix block}
        $G_{ij} \gets K_\otimes(Z_i, Z_j)$\;
                \BlankLine
        \tcc{Enforce unbiased estimator}
        \If{$i = j$}{
            $\text{diag}(G_{ij}) \gets 0$\;
        }        
        $Y_i \gets Y_i + G_{ij}  X_j$\;
    }
        $A \gets A + X_i^\top  Y_i$\;}
\BlankLine
$\text{KIPC} \gets \frac{1}{n - 1} \operatorname{Tr}(A)$\;
\Return $\text{KIPC}$\;
\end{algorithm}

~\newline


\bibliography{biblio}

\end{document}